\documentclass[a4paper,12pt]{article}
\usepackage{citeref}
\input cohmacros.sty

\def\at#1{[*** \att #1 ***]}  % attention
\def\at#1{} % switch off attention
\def\bfi#1{#1} % switch off bfi
\def\bfix#1{{\bf #1}} % keep bfix

\advance\textheight3.8cm        % for maximal amount of text per page
\advance\topmargin-2.0cm
\advance\textwidth3.6cm
\advance\evensidemargin-1.6cm
\advance\oddsidemargin-1.6cm

\def\E{\mathbf E}
\def\indexp#1{}
\def\indexq#1{}

\begin{document}

\vspace*{-2cm}
\begin{center}
%%%%%%%%%%%%%%%%%%%%%%%%%%%%%%%%%%%%%%%%%%%%%%%%%%%%%%%%%%%%%%%%%%%%%%%%
{\LARGE \bf The Born rule -- 100 years ago and today} \\[4mm]

\vspace{1cm}

\centerline{\sl {\large \bf Arnold Neumaier}}

\vspace{0.5cm}

\centerline{\sl Fakult\"at f\"ur Mathematik, Universit\"at Wien}
\centerline{\sl Oskar-Morgenstern-Platz 1, A-1090 Wien, Austria}
\centerline{\sl email: Arnold.Neumaier@univie.ac.at}
\centerline{\sl \url{https://arnold-neumaier.at}}

\end{center}

%\hfill Born/Born2025.tex

\hfill arXiv:2502.08545

\hfill updated March 20, 2025

\vspace{0.5cm}

\bigskip
%%%%%%%%%%%%%%%%%%%%%%%%%%%%%%%%%%%%%%%%%%%%%%%%%%%%%%%%%%%%%%%%%%%%%%%
%{\small
\bfi{Abstract.}
Details of the contents and the formulations of the Born rule changed 
considerably from its inception by Born in 1926 to the present day. 
This paper traces the early history of the Born rule 100 years ago, 
its generalization (essential for today's quantum optics and quantum 
information theory) to POVMs around 50 years ago, and a modern 
derivation from an intuitive definition of the notion of a quantum 
detector.
It is based to a large extent on little known results from the recent
books 'Coherent Quantum Physics' (2019) by A. Neumaier and 
'Algebraic Quantum Physics, Vol. 1' (2024) by A. Neumaier and D. Westra,

Also discussed is the extent to which the various forms of the Born 
rule have, like any other statement in physics, a restricted domain of 
validity, which leads to problems when applied outside this domain.

\bigskip

{\bf Keywords:} Born rule, probability in quantum mechanics,
statistical interpretation of quantum mechanics

{\bf MSC2020 classification: 81-03, 81P10, 81P15} 

\bigskip

This research received no funding.

{\bf Acknowledgment.} I recall with pleasure heated discussions with 
Francois Ziegler, several years ago, about the detailed history of the 
Born rule. I also want to thank Roger Balian for discussions related 
to this paper, especially about the condensed form (BR-C) of Born's 
rule and its use in the maximum entropy principle.

\newpage
%%%%%%%%%%%%%%%%%%%%%%%%%%%%%%%%%%%%%%%%%%%%%%%%%%%%%%%%%%%%%%%%%%%%%%%%
\tableofcontents % Inhalt hier

%\vspace*{2cm}

\parindent=0pt
\openup 2pt
\parskip 2ex plus 1pt minus 1pt

\newpage
%%%%%%%%%%%%%%%%%%%%%%%%%%%%%%%%%%%%%%%%%%%%%%%%%%%%%%%%%%%%%%%%%%%%%%%%
\section{Ten formulations of the Born rule}\label{s.BornForm}

In today's quantum mechanics textbooks, the Born rule, in one or the 
other of its many forms, is taken to be the axiomatic cornerstone of 
Max Born's statistical interpretation of quantum mechanics, conceived 
in 1926. At that time, an unspecific probabilistic interpretation of 
quantum mechanics had already been in the air.

For example, we read in \sca{Jeans} \cite[p.91]{Jea} (1924):
{\it ''Obviously quantum-dynamics must be intimately concerned with 
the probability of such jumps. [...] measure, in some way at present 
unknown, the probabilities of jumps in the velocity and perhaps 
also in the position of an electron which forms part of an atomic 
system.''} 
And in \sca{Bohr} et al. \cite[p.75]{BohrKS1924} (1924), we read:
{\it ''In addition, we assume that the occurrence of transition 
processes, both for the given atom itself and for the other atoms, 
with which it communicates, is linked with this mechanism through 
probability laws that are analogous to the laws of Einstein's theory 
for the transitions induced by external radiation between stationary 
states.''}
\\(German original:
{\it ''Ferner nehmen wir an, dass das Vorkommen von 
\"Ubergangsprozessen, sowohl f\"ur das gegebene Atom selbst wie f\"ur 
die anderen Atome, mit denen es kommuniziert, mit diesem Mechanismus 
durch Wahrscheinlichkeitsgesetze verkn\"upft ist, die den Gesetzen der 
Einsteinschen Theorie f\"ur die von \"ausserer Strahlung induzierten 
\"Uberg\"ange zwischen station\"aren Zust\"anden analog sind.''})

In February 1926, this interpretation was already semiquantitative, 
valid for large quantum numbers: 
{\it ''According to the quantum theory, the inner state of an atom can 
change only in sudden jumps. Only the discrete stationary states of 
the atoms are possible. In place of the classical continuous absorption 
of energy we therefore have a discontinuous one in single jumps, for 
which certain transition probabilities exist. Between both must then 
exist correspondence relations, i.e., an approximate, for large quantum 
numbers even exact, agreement between the classical and the quantum 
theoretical calculation.''}
\\(German original, \sca{Nordheim} \cite[p.512]{Nor1926}:
{\it ''Nach der Quantentheorie kann sich der innere Zustand eines Atoms
nur sprunghaft \"andern. Es sind nur die diskreten station\"aren 
Zust\"ande der Atome m\"oglich. An Stelle der klassischen 
kontinuierlichen Energieaufnahme haben wir daher eine 
diskontinuierliche in einzelnen Spr\"ungen, f\"ur die gewisse 
\"Ubergangswahrscheinlichkeiten bestehen. Es m\"ussen dann zwischen 
beiden korrespondenzm\"assige Beziehungen vorhanden sein,
d.h. n\"aherungsweise, f\"ur grosse Quantenzahlen sogar exakt, 
\"Ubereinstimmung zwischen der klassischen und quantentheoretischen 
Rechnung herrschen.''}) 

At that time, communication by letters, preprints, travel, and 
conferences played an important role, and novelties were not always 
reflected in the publication date. Given the Schr\"odinger equation 
(interpreted by \sca{Schr\"odinger} \cite{Schr1926} as a charge 
density), the nontrivial part of the statistical interpretation was to 
show that the statistical properties of scattering experiments were 
consistent with, and could be derived from the deterministic 
Schr\"odinger equation, 

In June 1926, \sca{Born} \cite[p.822]{Bor1926b} found a highly 
successful statistical interpretation of the then very new wave 
function of Schr\"odinger, from which he could derive very specific 
physical conclusions, namely the qualitative features of collision 
experiments (which were later verified quantitatively). 
This earned Born the Nobel prize in 1954 {\it ''for his fundamental 
research in quantum mechanics, especially for his statistical 
interpretation of the wavefunction''} \cite{Nobel1954}.

Wallner's Nobel prize presentation speech for the general public 
comments (\cite{Nobel1954}):
{\it ''Schr\"o\-dinger had not solved the problem of how it is possible 
to make statements about the positions and velocities of particles if 
one knows the wave corresponding to the particle.
Born provided the solution to the problem. He found that the waves 
determine the probability of the measuring results. For this reason, 
according to Born, quantum mechanics gives only a statistical 
description.''} 
This is a bit inaccurate, as Born found probabilities for state 
transitions of particles, but not probabilities for positions and 
velocities of particles -- this was done a little later by Pauli.
This motivates a detailed study of the history of the statistical 
interpretation of quantum mechanics.

This paper traces the early history of the Born rule 100 years ago, 
its generalization (essential for today's quantum optics and quantum 
information theory) to POVMs 50 years ago, and a modern derivation 
from an intuitive definition of the notion of a quantum detector.
It is based to a large extent on little known results from my recent
books (\sca{Neumaier} \cite{Neu.CQP} and \sca{Neumaier \& Westra} 
\cite{NeuW}) and three unpublished preprints (\sca{Neumaier} 
\cite{Neu.BornM,Neu.IFound,Neu.tomo}).

%%%%%%%%%%%%%%%%%%%%%%%%%%%%%%%%%%%%%%
\subsection{The Born rule before 1970}

The Born rule has many slightly inequivalent formulations, ten 
different forms of which we state in the following, in modern 
terminology. 
Two formulations, the universal form (BR-US) and the discrete form 
(BR-DS) are taken almost verbatim from Wikipedia \cite{Wik.Born}.
Four other formulations. are taken with minor changes from Chapter 14 
of \sca{Neumaier} \cite{Neu.CQP} (but with the scattering form restated 
in two almost identical formulations with different meaning), and one 
more from Part II of \sca{Neumaier \& Westra} \cite{NeuW}. Two more 
general formulations were discovered only after 1970, and are described 
in Subsection \ref{s.POVM}. Finally, a tenth form (BR-Q) is presented 
in Subsection \ref{s.qEx}, adapted to quantum statistical mechanics
and quantum field theory.

\bfix{(BR-OSc), Born rule (objective scattering form)}:
In a scattering experiment described by the S-matrix $S$,
\lbeq{e.psiS}
\Pr(\psi_\out|\psi_\iin):=|\psi_\out^*S\psi_\iin|^2
\eeq
is the conditional probability density that scattering of particles
prepared in the in-state $\psi_\iin$ results in particles
in the out-state $\psi_\out$. Here the in- and out-states
are asymptotic eigenstates of the Hamiltonian, labelled by a maximal
collection of independent quantum numbers (including the energies, 
momenta and angular momenta of the particles involved).

\bfix{(BR-OE), Born rule (objective expectation form)}:
The value of a quantity corresponding to an operator $X$ of a system 
in the pure state $\psi$ or the mixed state $\rho$ equals on average 
the quantum expectation value 
\lbeq{e.psiX}
\<X\>:=\psi^*X\psi,
\eeq
or
\lbeq{e.rhoX}
\<X\>:=\tr\rho X,
\eeq
respctively. Here $\tr$ denotes the trace of an operator.

\bfix{(BR-MSc), Born rule (measured scattering form)}:
In a scattering experiment described by the S-matrix $S$, formula 
\gzit{e.psiS} gives the conditional probability density that when 
particles prepared in the in-state $\psi_\iin$ are scattered, they are 
found in the out-state $\psi_\out$. Here the in- and out-states
are asymptotic eigenstates of the Hamiltonian, labelled by a maximal
collection of independent quantum numbers (including the energies, 
momenta and angular momenta of the particles involved).

As part of the Born rule, it is frequently but not always stated that 
the results of the measurement of a quantity exactly equals one of the 
eigenvalues.

\bfix{(BR-US), Born rule (universal spectral form)}:
If a quantity corresponding to a self-adjoint Hermitian operator
$X$ is measured in a system described by a pure state with normalized
wave function $\psi$ then
\\
(i) the measured result will be one of the eigenvalues $\lambda$ of
$X$, and
\\
(ii) for any open interval $\Lambda$ of real numbers, the probability
of measuring $\lambda\in\Lambda$ equals $\psi^*P(\Lambda)\psi$, where
$P(\Lambda)$ is the projection onto the invariant subspace
of $X$ corresponding to the spectrum in $\Lambda$ by the spectral 
theorem.

Since the formulation involves projection operators, measurements 
satisfying (BR-US) are called \bfix{projective measurements};
other name for these are \bfi{von Neumann measurements} and
\bfi{ideal measurements}. Not all measurements are of this kind; 
see Subsection \ref{s.POVM} for a more general class of measurements,
and Subsection \ref{s.valid} for a fuller discussion of the practical
limitations the class of projective measurements has. 

\bfix{(BR-FS), Born rule (finite spectral form)}: 
In a projective measurement of a Hermitian, self-adjoint operator $X$
(or vector of pairwise commuting operators)  
with a finite spectrum, the possible values are precisely the finitely 
many eigenvalues $\lambda_k$ of $X$ (or joint eigenvalues of the 
components of $X$), measured with a probability of 
\lbeq{e.rhoP}
p_k=\tr \rho P_k,
\eeq
where $P_k$ is the orthogonal projector to the eigenspace of $X$ 
corresponding to $\lambda_k$.

\bfix{(BR-DS), Born rule (discrete spectral form)}:
Suppose that a quantity corresponding to a self-adjoint Hermitian 
operator $X$ with a discrete spectrum is measured in a system described 
by a pure state with normalized wave function $\psi$. Then:
\\
(i) the measured result will be one of the eigenvalues $\lambda$ of
$X$, and
\\
(ii) the probability of measuring a given eigenvalue $\lambda$ equals
$\psi^*P\psi$, where $P$ is the projection onto the eigenspace
of $X$ corresponding to $\lambda$.

\bfix{(BR-ME), Born rule (measured expectation form)}:
If a quantity corresponding to a self-adjoint Hermitian operator 
$X$ is measured on a system in the pure state $\psi$ or the mixed 
state $\rho$, the results equal on average the quantum expectation 
value \gzit{e.psiX} or \gzit{e.rhoX}, respectively.

%%%%%%%%%%%%%%%%%%%%%%%%%%%%%%%%%%%%%
\subsection{The Born rule after 1970}\label{s.POVM}

Nearly 50 years after Born, a theoretical description of more general 
quantum measurements was introduced in 1970 by \sca{Davies \& Lewis} 
\cite{DavL}. A much more readable account was given by 
\sca{Ali \& Emch} \cite{AliE.meas} in terms of 
\bfix{positive operator valued measures} (\bfix{POVM}s). 

These general measurement schemes are based on the concept of a
\bfix{finite quantum measure} (also called a 
\bfix{finite resolution of unity}),\index{resolution of unity!finite}
a family of finitely many Hermitian positive semidefinite operators 
$P_k$ on a Hilbert space summing to 1,
\[
\sum_k P_k=1.
\]
This straightforward generalization of the concept of a finite
probability measure, given by a family of finitely many nonnegative
numbers $p_k$ (probabilities) summing to 1, is a simplified version of
a finite POVM.
The following generalization of the finite spectral form (BR-FS) 
of the Born rule is essentially in 
\sca{Ali \& Emch} \cite[(2.14)]{AliE.meas}.

\bfix{(BR-POVM), Born rule (POVM form)}: 
In a general quantum measurement, the possible values are finitely 
many distinct numbers or vectors $\lambda_k$, measured with a 
probability of \gzit{e.rhoP}, where the $P_k$ form a quantum measure.

Progressing from the Born rule to so-called positive operator valued
measures (POVMs) is already a big improvement, commonly used in quantum
optics and quantum information theory. 
Many measurements in quantum optics are POVM measurements
\cite{wik.POVM}, i.e., described by a positive operator-valued measure.
These follow a different law of which the Born rule is just a very 
special case where the POVM is actually projection-valued. 
Quantum measures are indispensable in quantum information theory
(\sca{Nielsen \& Chuang} \cite{NieC}).
They are able to account for things like losses, 
imperfect measurements, limited detection accuracy, dark detector 
counts, and the simultaneous measurement of position and momentum. 

Quantum measures are also needed to describe quite ordinary experiments 
without making the traditional textbook idealizations; see, e.g., 
\sca{Busch} et al. \cite{BusGL}.
For a fairly concise, POVM-based exposition of the foundations of 
quantum mechanics see, e.g., \sca{Englert} \cite{Eng}. A short history 
can be found in \sca{Brandt} \cite{Bra}.

The quantum measure description of a real measurement device cannot 
simply be postulated to consist of orthogonal projectors. The correct
quantum measure must be found out by quantum measurement tomography, 
guided by the theoretical model of the measuring equipment, then 
ultimately calibrating it using the formula (BR-POVM). 
See, e.g., \sca{D'Ariano} et al. \cite{DArPS} and 
\sca{Neumaier} \cite{Neu.tomo}.

To justify (BR-POVM), the usual practice is to assume the Born rule 
for projective measurements as a basic premise, with a purely 
historical justification. Later (if at all), the more general quantum 
measure (POVM) setting is postulated and justified (e.g., in
\sca{Fuchs \& Peres} \cite{FucP}) in terms of the Born rule for 
projective measurements, in an extended state space formally 
constructed using an appropriate ancilla on the basis of Naimark's 
theorem (\sca{Naimark} \cite{Nai}). This shows the consistency with 
the Born rule, but the ancilla has not always a meaning in terms 
of the physical Hilbert space.

\bigskip 

Following \sca{Neumaier \& Westra} \cite[Part II]{NeuW}, we give in
Section \ref{s.today} precise definitions that lead to the 
correspondence between statistical expectations and quantum 
expectation claimed by the following condensed form of the Born rule. 

\bfix{(BR-C), Born rule (condensed form)}: 
{\it The statistical expectation of the measurement results equals the
quantum expectation of the measured quantity.}

Both (BR-POVM) and (BR-C) are derived in Subsection \ref{s.derived} 
from a new, nontechnical measurement principle that is easy to motivate 
and understand, without any technicalities regarding the spectrum of 
operators.

%%%%%%%%%%%%%%%%%%%%%%%%%%%%%%%%%%%%%%%
\subsection{Quantum expectation values}\label{s.qEx}

The quantum expectation values \gzit{e.psiX} and \gzit{e.rhoX} are 
purely mathematical definitions, hence belong to the core of 
uninterpreted quantum mechanics. Their use in calculations is 
completely independent of any (assumed or questioned) relations of the
quantum mechanical formalism to reality. The latter enters only through 
some version of the Born rule. 

In quantum statistical mechanics and quantum field theory one usually 
makes a theoretical analysis of quantum processes where no actual 
measurement is made before the process is completed. Since one cannot 
invoke formulations of the Born rule that make reference to measurement,
all probabilities and expectation values that appear in such an 
analysis, must be considered as formal expressions 
\gzit{e.psiX}--\gzit{e.rhoP}, with a clear mathematical meaning 
independent of the interpretation of quantum mechanics in terms of
reality, knowledge, or measurement. 

Every knowledge-based approach to statistical mechanics needs (BR-C) at 
the very basis, though this is usually left implicit and only tacitly 
assumed. Indeed, the only way to know something about quantum 
expectation values in an assumed model of a real quantum system is to 
gather information about the system or its source, and to translate it 
into information about the state of the system. The latter is usually 
done by means of the \bfix{maximum entropy principle}, which 
(\sca{Balian \& Balazs} \cite{BalB}) directly equates the statistical 
expectation value with the quantum expectation value in order to be 
able to estimate the state.
Note that even Heisenberg's uncertainty relation is a statement about
(theoretical) quantum expectation values that needs (BR-C) to imply 
anything about (measured) statistical uncertainty.
  
Since (BR-OSc) and (BR-OE) do not refer to measurement, they are not 
loaded with the difficulties surrounding the measurement problem of 
quantum mechanics. However, as we shall discuss in more detail in 
Subsection \ref{s.paradox}, were abandoned already by 1930, being 
considered as mathematically inconsistent.
(A referee remarked that Bohmian mechanics restored the consistency of 
an objective, measurement-free view. But the restoration in Bohmian 
mechanics was only partial, namely for position, whereas there is still 
no objective interpretation of the measurement of energy and angular 
momentum. Thus (BR-OSc) and (BR-OE) do not hold.)

In particular, without objective interpretation, the quantum 
probabilities and quantum expectation values are purely theoretical 
quantities without any direct relation to reality. The link to reality 
is exclusively in the subject possessing the knowledge considered. 

If one nevertheless wants to give the computations in quantum 
statistical mechanics an objective physical meaning, one must therefore 
define the objective quantum value of a vector $X$ of operators to 
{\it be} the quantum expectation value $\<X\>$:

\bfix{(BR-Q), Born rule (quantum value form)}:
The quantum value $\ol X$ of a quantity corresponding to an operator 
$X$ of a system in the pure state $\psi$ or the mixed state $\rho$ is 
defined as the quantum expectation value \gzit{e.psiX} or \gzit{e.rhoX},
respectively. It is approximately measured, with an uncertainty of at
least
\lbeq{e.sigmaX}
\sigma_X:=\sqrt{\<(X-\ol X)^*(X-\ol X)\>}.
\eeq
Here $X^*$ is the adjoint of $X$.

This form of the Born rule expresses the spirit of Heisenberg's 
statement (in his 1927 paper on the uncertainty relation):
{\it ''To every quantum theoretical quantity or matrix, one may 
associate a number, that gives its 'value', with a certain probable 
error; the probable error depends on the coordinate system; for every
quantum theoretical quantity there is a  coordinate system in which 
the probable error for this quantity vanishes.''}
\\(German original, \sca{Heisenberg} \cite[p.181f]{Hei1927}:
{\it '' Jeder quantentheoretischen Gr\"osse oder Matrix l\"asst sich 
eine Zahl, die ihren 'Wert' angibt, mit einem bestimmten 
wahrscheinlichen Fehler zuordnen; der wahrscheinliche Fehler h\"angt 
vom Koordinatensystem ab; f\"ur jede quantentheoretische Gr\"osse gibt 
es je ein Koordinatensystem, in dem der wahrscheinliche Fehler f\"ur 
diese Gr\"osse verschwindet.''}) 
(\sca{Heisenberg} \cite[p.181f]{Hei1927})

The quantum value form (BR-Q) of the Born rule is the basis of the 
\bfi{thermal interpretation of quantum mechanics}, discussed in my 
book \cite{Neu.CQP}. 

In quantum field theory, quantum expectation values of non-Hermitian 
operators are routinely used; they are already needed in the definition 
of so-called $N$-point functions as quantum expectation values of 
products of field operators. That $2$-point functions are in principle 
observables through linear response theory shows that the operators 
corresponding to observable quantities need not be Hermitian. 
Indeed, Dirac's 1930 quantum mechanics textbook 
(\sca{Dirac} \cite[p.24]{Dir1}), which introduced the name  
''observable'' for operators, used this terminology for arbitrary 
linear operators: {\it ''It is convenient to count any operator that 
can be multiplied into the $\psi$s and $\phi$s in accordance with the 
foregoing axioms as an observable.''}

It is therefore interesting to note that all later editions of Dirac's 
textbook and {\it all} later textbooks on quantum mechanics require 
the restriction to Hermitian operators possessing a real spectral 
resolution, i.e., in modern terminology, to self-adjoint Hermitian 
operators. Probably Dirac became aware of the fact that when $X$ is 
not normal (and in particular when $X$ is defective, hence has not even 
a spectral resolution), it is impossible to give the quantum value 
$\<X\>$ of $X$ a spectral interpretation -- which before 1970 was an 
essential ingredient of all measurement-based forms of the Born rule. 
Only the more recent formulations (BR-POVM), (BR-C), and (BR-Q) are 
again free of a spectral restriction.

%%%%%%%%%%%%%%%%%%%%%%%
\subsection{Discussion}

The mathematically consistent relaxation (BR-Q) of (BR-OE) gives up the 
assumption 

(D) If the value of $X$ is $x$ then the value of $f(X)$ is $f(x)$,

tacitly made in classical statistical physics, and imported from there 
into the spectral forms of the Born rule. In practice, we only measure 
a small set of distinguished variables suc has distances, position, 
energy, momentum, angular momentum, spin, helicity, and not arbitrary 
nonlinear expressions in these. Hence it is quite reasonable to require 
(D) not for arbitrary nonlinear $f$ but only for linear $f$. 

Note that giving up (D) is necessary if one wants to account for 
experiments involving the simultaneous measurement of position and 
momentum, such as that routinely used in modern collision experiments, 
where particle tracks with fairly well-determined positions and momenta 
are reconstructed in in a time projection chamber. 
A POVM representation of the latter is given in 
\sca{Neumaier \& Westra} \cite[Section 6.2.4]{NeuW}. Indeed, (D) is 
also incompatible with (BR-POVM). 

The three newest forms of the Born rule, (BR-POVM), (BR-C) , and (BR-Q),
are mathematically much more elementary since they do not depend on the 
notion of self-adjointness of operators. The POVM form (BR-POVM) of the 
Born rule is more or less equivalent to the condensed form (BR-C);
see Subsection \ref{s.derived}. As already stated in Subsection 
\ref{s.POVM}, (BR-POVM) is a generalization of the finite spectral form 
(BR-FS). 

\bigskip

The measured expectation form (BR-ME) of the Born rule asserts that 
measurements result in a random variable whose expectation agrees with 
the formal quantum expectation. The average in question cannot be taken 
as a sample average (where only an approximate equal results, with an 
accuraccy depending on size and independence of the sample) but must be 
considered as the theoretical expectation value of the random variable.
Thus (BR-ME) implies (BR-C). On the other hand, (BR-C) is more general 
since it makes no self-adjointness assumption. For example, (BR-C)
applies to complex measurements of the operator $A=p+iq$, equivalently
to the joint measurement of position $q$ and momentum $p$.
(By Heisenberg's uncertainty relation, these cannot be measured jointly 
with arbitrary accuracy, but projective measurements do not allow them 
to be jointly measured at all.) 

Mathematically, the expectation value of a random variable is completely
insensitive to the results of a finite number of realizations, just as 
the limit of a sequence does not change when finitely many sequence 
entries are changed arbitrarily. Since we can only take finitely many 
measurements on a system, the measured expectation form (BR-ME) of the 
Born rule says, strictly speaking, nothing at all about individual 
measurement results.

If the measurement result $\lambda_i$ is an isolated eigenvalue of $A$,
the universal form (BR-US) reduces to the discrete form (BR-DS), since 
one can take $\Lambda$ to be an open interval intersecting the spectrum 
in $\lambda_i$ only, and in this case, $P(\Lambda)=P_i$. The discrete
form (BR-DS) clearly implies the finite form (BR-FS) for the case where 
$A$ is a scalar. (BR-FS) takes into account that realistic measurements 
can distinguish only finitely many measurement results.
(BR-FS) also allows for the joint measurement of commuting 
quantities, but similar extensions could be formulated for the discrete 
form and the universal form. For projective measurements, (BR-FS) is 
derived from (BR-C) in Section \ref{s.derived}.

If, in a projective measurement with 1-dimensional projectors 
$P_k=\phi_k\phi_k^*$, the source is pure, described by 
$\rho=\psi\psi^*$ with
the normalized state vector $\psi\in\Hz$, then \gzit{e.rhoP} can be 
written in the more familiar squared amplitude form
\lbeq{e.BornSquare}
p_k=|\phi_k^*\psi|^2.
\eeq
Together with the relation $\psi=S\psi_\iin$, which links $\psi$ to the 
asymptotic in-state via the S-matrix, this shows that the finite 
spectral form (BR-FS) implies the measured scattering form (BR-MSc). 

Using the spectral theorem, it is not difficult to show that the
universal form of the Born rule implies the measured expectation form.
Conversely, when one assumes von Neumann's axiom D, which effectively 
forces the measurements to be projective measurements, the measured 
expectation form (BR-ME) of the Born rule implies the second part (ii)
of the universal form (BR-US). It follows that the first part (i) holds 
with probability 1, but not with certainty, as the universal form 
requires.
This is not just hair splitting. The difference between probability 1
and certainty can be seen by noting that a random number drawn 
uniformly from $[0,1]$ is irrational with probability 1,
while measurements usually produce rational numbers.

\sca{von Neumann} \cite[p.255]{vNeu1927b} derives for a 
quantum expectation value satisfying four (for von Neumann) plausible
conditions (conditions A.--D. listed and motivated on p.250--252) the 
necessity of the formula \gzit{e.rhoX} with a Hermitian density 
operator $\rho$ (his $U$) of trace 1. This is abstract mathematical
reasoning, still within the core of formal (uninterpreted) quantum 
mechanics, independent of any relation to measurement.
Note that assuming von Neumann's condition D. which is a restricted 
form of condition (D) above, forces the measurements to be projective 
measurements. Thus he misses the POVM generalization of the Born rule 
since the latter violates his condition D.

\newpage
%%%%%%%%%%%%%%%%%%%%%%%%%%%%%%%%%%%%%%%%%%%%%%%%%%%%%%%%%%%%%%%%%%%%%%%%
\section{The Born rule 100 years ago}

%%%%%%%%%%%%%%%%%%%%%%%%%%%%%%%%%%%%%%%%%
\subsection{The genesis of the Born rule}\label{s.BornGenesis}

The term 'Born rule' appeared quite late, apparently first in 1934 by 
\sca{Bauer} \cite[p.302]{Bau34} ({\it ''la r\`egle de Born''}; 
cf. \sca{Ziegler} \cite{Zie}). Before that, during the gestation
period of finding the right level of generalization and interpretation,
the pioneers talked more vaguely about Born's interpretation of quantum
mechanics (or of the wave function). For example, in 1927, \sca{Jordan}
\cite[p.811]{Jor1927} wrote about 
{\it ''Born's interpretation of the solution [of the] Schr\"odinger 
equation''}
\\(German original:
{\it ''Born's Deutung der L\"osung [der] Schr\"odingergleichung''}). 

\sca{Wessels} \cite[p.187]{Wes} even goes so far as to claim that
{\it ''Neither Born nor most of his contemporaries saw in it anything 
that even suggests the interpretation that we now associate with his 
name.''} Thus it is interesting to consider the genesis of the Born 
rule, based on the early papers of the pioneers of quantum mechanics.
\sca{Mehra \& Rechenberg} \cite{MehR} give a detailed history of 
Born's statistical interpretation, and \sca{Bacciagaluppi} 
\cite{Bac2020} expands on the early years 1926--1927. However, they 
miss a number of subtle points that shed new light on the early 
developments.

Born originally related his interpretation not to measurement but 
to objective properties of scattering processes, no matter whether or 
not these were observed.

The two 1926 papers by \sca{Born} \cite{Bor1926a,Bor1926b} (the first
being a summary of the second) introduced the probabilistic
interpretation.  His 1926 formulation
{\it ''determines the probability that the electron cominf from the 
$z$-direction is thrown into the direction determined by 
$\alpha,\beta,\gamma$ (with a phase change of $\delta$)''}
\\(German original, \sca{Born} \cite[p.865f]{Bor1926a}: 
{\it ''bestimmt die Wahrscheinlichkeit daf\"ur, dass das aus der
$z$-Richtung kommende Elektron in die durch $\alpha,\beta,\gamma$
bestimmte Richtung (und mit einer Phasen\"anderung $\delta$) geworfen
wird''})
when rephrased in modern terminology, is a special case of the 
objective scattering form (BR-OSc) of the Born rule. 
Note that Born didn't have the concept of an S-matrix, first introduced 
in 1937 by \sca{Wheeler} \cite{Whe1937}. 
(The S-matrix elements $S_{ik}$ are the inner products $e_i^*Se_k$, 
where the $e_i$ are the standard basis vectors of the matrix $S$; 
in Born's case the stationary states of the particle.
The relation with the S-matrix is visible more explicitly in 
\sca{Dirac} \cite{Dirac1927b}, who computed in 1927 what are today 
called the S-matrix elements in the so-called Born approximation.)
However, the square of the absolute value the S-matrix elements for 
single-particle scattering is proportional to the 
{\it ''Ausbeutefunktion''} $\Phi_{n,m}$ mentioned in  
\sca{Born} \cite[p.824]{Bor1926b}, already used (with similar notation) 
in \sca{Nordheim} \cite[p.512]{Nor1926}. 

\sca{Bacciagaluppi} \cite{Bac2020} comments:
{\it ''Importantly, the corresponding probabilities are not 
probabilities for 'finding' a system in a certain stationary state 
upon measurement: the atom and the electron (at least when the 
interaction is completed) are assumed to be in a stationary state.''}
Thus Born wrote about objective properties of electrons 
(''being thrown out'') independent of measurement. His statement does 
not depend on anything being measured, let alone to assigning a
precise numerical measurement value! 

The 1927 paper by \sca{Born} \cite{Bor1927} extends this rule on p.173
to probabilities for quantum jumps (''Quantensprung'', p.172)
between energy eigenstates, given by the absolute squares of inner
products of the corresponding eigenstates, still using objective rather
than measure\-ment-based language: For a system initially in state $n$ 
given by formula (9) in Born's paper he says,
{\it ''The square $|b_{nm}|^2$ is, according to our basic hypothesis,
the probability that after the end of the disturbance, the system is in
state $m$''}
\\(German original: 
{\it ''Das Quadrat $|b_{nm}|^2$ ist gem\"ass unserer Grundhypothese
die Wahrscheinlichkeit daf\"ur, dass das System sich nach Ablauf der
St\"orung im Zustand $m$ befindet''})
Here state $n$ is the $n$th stationary state (eigenstate with a
time-dependent harmonic phase) of the Hamiltonian.

Born derives this rule from two assumptions. The first assumption, made
on p.170 and repeated on p.171 after (5), is that an atomic system is
always in a definite stationary state:
{\it ''Thus we shall stick to Bohr's picture that an atomic system is 
always only in a stationary state. [...] but in general we will know 
in any moment only that, because of the prior history and the given 
physical conditions, there is a certain probability that the atom is 
in the $n$th state.''}
\\(German original: 
{\it ''Wir werden also an dem Bohrschen Bilde festhalten, dass ein
atomares System stets nur in einem stationaren Zustand ist. [...]
im allgemeinen aber werden wir in einem Augenblick nur wissen,
dass auf Grund der Vorgeschichte und der bestehenden physikalischen
Bedingungen eine gewisse Wahrscheinlichkeit daf\"ur besteht, dass das
Atom im $n$-ten Zustand ist.''})

Thus for the early Born, the quantum numbers of the (proper or improper)
stationary states of the system are objective properties of a quantum 
mechanical system. This assumption works indeed for equilibrium quantum
statistical mechanics -- where expectations are defined in terms of the
partition function and a probability distribution over the stationary
states. It also works for nondegenerate quantum scattering theory
-- where only asymptotic states figure. However, it has problems in the
presence of degeneracy, where only the eigenspaces, but not the
stationary states themselves, have well-defined quantum numbers.
Indeed, Born assumes -- on p.159, remark after his (2) and
his Footnote 2 -- that the Hamiltonian has a nondegenerate, discrete
spectrum.

Born's second assumption is his basic hypothesis on p.171 for
probabilities for being (objectively) in a stationary state:
{\it ''[...] is a certain probability that the atom is in the $n$th 
state. We now claim that as measure for this state probability, one 
has to choose the quantity $|c_n|^2=|\int \psi(x,t)\psi_n^*(x)dx|^2$.''}
\\(German original: 
{\it ''[...] eine gewisse Wahrscheinlichkeit daf\"ur besteht, dass
das Atom im $n$-ten Zustand ist. Wir behaupten nun, dass als Mass
dieser Zustandswahr\-scheinlichkeit die Gr\"osse
$|c_n|^2=|\int \psi(x,t)\psi_n^*(x)dx|^2$ zu w\"ahlen ist.''})
This (now obsolete) interpretation of objective stationary states 
persisted for some time in the literature; e.g., in the 1929 book by 
\sca{de Broglie} \cite{dBro1929}. In the introduction he refers both 
to an objective position-based and to a measurement-based spectral 
formulation:

p.4: 
{\it ''In the proper domain of the new dynamics, one can usually rely 
on the principle that the square of the wave amplitude, the intensity,
at a particular place and a particular time gives the probability 
that the particle is at this time at this place. One ckecks easily that 
this principle is necessary to explain diffraction and interference.''}
\\(German original: 
{\it ''Im eigentlichen Gebiet der neuen Dynamik kann man sich meistens 
auf das Prinzip verlassen, dass das Quadrat der Wellenamplitude, die 
Intensit\"at, an einem bestimmten Ort und zu einer bestimmten Zeit die 
Wahrscheinlichkeit daf\"ur liefert, dass sich das Teilchen zu dieser 
Zeit an diesem Ort befindet. Man \"uberlegt leicht, dass dieses Prinzip 
notwendig ist, um Beugung und Interferenz zu erkl\"aren.''})

p.6f, after mentioning difficulties of the objective position-based 
view: 
{\it ''Finally, there is still a fourth approach, which currently has 
the most supporters. It was developed by Heisenberg and Bohr. [...]
According to this view , the wave [...] is only a symbolic 
representation of what we know about the particle. [...] Thus the 
intensity distribution and spectral nature of the wave allws one to 
give the probability that an experiment performed at time $t$ finds 
the particle at a particular place, or assigns to it a particular 
state of motion.''}
\\(German original: 
{\it ''Schliesslich gibt es noch eine vierte Betrachtungsweise, die
augenblicklich die meisten Anh\"anger hat. Sie wurde von Heisenberg und
Bohr entwickelt. [...] Nach dieser Auffassung ist die Welle [...] nur
eine symbolisch Darstellung von dem, was wir \"uber das Teilchen wissen.
[...] so gestatten Intensit\"atsverteilung und Spektralbeschaffenheit
der Welle, die Wahrscheinlichkeit daf\"ur anzugeben, dass ein zur
Zeit $t$ ausgef\"uhrtes Experiment das Teilchen an einem bestimmten
Ort findet oder ihm einen bestimmten Bewegungszustand zuschreibt.''})

The objective interpretation is needed to be able to argue on p.70 as
follows:
{\it ''The density of the ensemble can also be regarded as the 
probability that a particle, whose motion belongs to the class 
considered, is at a particular time at a particular point, if its exact 
position is unknown. [...] We had denoted this assumption in the 
introduction as 'interference principle'.''}
\\(German original: 
{\it ''Die Dichte des Schwarms kann man auch als die Wahrscheinlichkeit 
daf\"ur betrachten, dass ein Teilchen, dessen Bewegung zur betrachteten 
Klasse geh\"ort, sich zu einer bestimmten Zeit an einem bestimmten 
Punkt befindet, wenn seine genaue Lage unbekannt ist. [...] Wir hatten 
diese Annahme in der Einleitung als 'Interferenzprinzip' bezeichnet.''})
However, de Broglie mentions at several places the associated 
difficulties.

In the main text, he mentions on p.117f on the one hand 
$a^2=|\psi(x)|^2$ ({\it ''by the interference principle''} of 
Schr\"odinger, pp.3--5), and on the other hand $a_k^2$ 
({\it ''by the postulate of Born''}), now in an objective formulation:
{\it ''When the wave mechanics was still at the beginning of its 
development, Max Born already made the proposal to regard every 
quantity $a_k^2$ as the relative probability for a particle to have 
the state of motion belonging to $\psi$. [...] We call this postulate 
of Born the 'principle of spectral resolution'. Assuming its validity, 
the determination of the particle through the corresponding wave has a 
double uncertainty: On the one hand, the position of the particle is 
undetermined, since by the interference principle, there is a positive 
probability that one finds the particle in any place of the spatial 
region occupied by the wave packet, being equal to the resulting 
intensity $a^2$. On the other hand, the state of motion, measured 
through momentum and energy, is by the principle of spectral resolution 
also undetermined, since there are several possible states of motion,
and the probability for each one is the amplitude square of the 
corresponding monochromatic component in the spectral resolution of 
the wave packet.''}
\\(German original: 
{\it ''Als die Wellenmechanik noch am Anfang ihrer Entwicklung stand,
hat Max Born schon den Vorschlag gemacht, jede Gr\"osse $a_k^2$ als
die relative Wahrscheinlichkeit daf\"ur aufzufassen, dass das
Teilchen den zu $\psi$ geh\"origen Bewegungszustand hat. [...]
Wir nennen dieses Bornsche Postulat das 'Prinzip der spektralen
Zerlegung'. Nimmt man seine G\"ultigkeit an, so hat die Bestimmung des
Teilchens durch die zugeh\"orige Welle eine doppelte Unbestimmtheit:
einerseits ist die Lage des Teilchens unbestimmt, da es nach dem
Interferenzprinzip eine endliche Wahrscheinlichkeit gibt, dass man das
Teilchen an irgendeiner Stelle des Raumgebiets findet, das vom
Wellenzug ausgef\"ullt wird, und zwar ist sie gleich der resultierenden
Intensit\"at $a^2$. Andererseits ist der durch Impuls und Energie
gemessene Bewegungszustand nach dem Prinzip der spektralen Zerlegung
ebenfalls unbestimmt, weil es mehrere m\"ogliche Bewegungszust\"ande
gibt und die Wahrscheinlichkeit f\"ur jeden durch das Amplitudenquadrat
der zugeh\"origen monochromatischen Komponente in der spektralen
Zerlegung des Wellenzugs ist.''})

De Broglie uses here exactly the same terminology as Born did in his
1927 paper \cite[p.168]{Born1927a},
{\it ''The particles are always accompanied by a wave process; these
de Broglie--Schr\"odinger waves depend on the forces and determine 
through the square of their amplitude the probability that a particle 
is present in the state of motion corresponding to the wave.
In hist last communication, Schr\"odinger also discussed the amplitude 
square of the waves and introduces for it the term 'weight function', 
which is already very close to the terminology of statistics. ''}
\\(German original: 
{\it ''Die Teilchen sind immer von einem Wellenvorgang begleitet; 
diese de Broglie--Schr\"odingerschen Wellen h\"angen von den Kr\"aften 
ab und bestimmen durch das Quadrat ihrer Amplitude die 
Wahr\-scheinlichkeit daf\"ur, dass ein Teilchen in dem der Welle 
entsprechenden Bewegungszustand vor\-handen ist.
In seiner letzten Mitteilung besch\"aftigt sich auch Schr\"odinger mit 
dem Amplitudenquadrat der Wellen und f\"uhrt daf\"ur den Ausdruck 
'Gewichtsfunktion' ein, der sich der Terminologie der Statistik schon 
sehr n\"ahert.''})
Thus we know that {\it ''Bewegungszustand''} (''state of motion'')
stands for the energy-momentum eigenstate of the particle. Indeed, 
''state of motion'' was at that time a technical term now no longer 
in use. The 1898 book by \sca{Jordan} \cite[p.1f]{Jor1898} defines:
{\it ''With the word 'force', one denotes the cause of a change of the 
state of motion of a body. -- As state of motion one has to understand 
not only a motion of any kind but also the rest -- as lack of any 
motion (velocity = 0, cf. p.3). [...] Every body maintains without 
change the state of motion that it has at any motion, in direction and 
speed, as long as no external force affects a change.''}
\\(German original: 
{\it ''Mit dem Worte Kraft bezeichnet man die Ursache einer \"Anderung 
des Bewegungszustandes eines K\"orpers. -- Als Bewegungszustand ist 
nicht nur eine Bewegung irgend welcher Art, sondern auch die Ruhe -- 
als Abwesenheit jeglicher Bewegung (Geschwindigkeit = 0, vgl. S. 3) --
aufzufassen. [...] Jeder K\"orper beh\"alt den Bewegungszustand, den er 
in irgend einem Momente hat, nach Richtung und Geschwindigkeit 
unver\"andert bei, so lange keine \"aussere Kraft (\"andernd) auf ihn 
einwirkt.''})
Thus Bewegungszustand is a state that persists unless 
changed by an interaction. 
In 1925, \sca{Kramers \& Heisenberg} \cite[p.692]{KraH} 
use the term in the quantum context for stationary states:
{\it ''which refer to two states of motion for which the
values of the quantities $J_1 \cdots J_s$ differ by 
$\tau_1h\cdots\tau_sh$. [...] We represent the states of motion by
points in a $J_1, J_2$-plane, the drawing plane of the figure.''}
\\(German original:
{\it ''die sich auf zwei Bewegungszust\"ande beziehen, f\"ur die die 
Werte der Gr\"ossen $J_1 \cdots J_s$ sich um $\tau_1h\cdots\tau_sh$ 
unterscheiden. [...] die Bewegungszust\"ande seien durch Punkte
in einer $J_1, J_2$-Ebene, der Zeichenebene der Figur, dargestellt.''})

%%%%%%%%%%%%%%%%%%%%%%%%%%%%%%%%%%%%%%%%%%%%%%%%%%%%%%%%%%%%%%%%%%
\subsection{Early objective, measurement-independent formulations}
\label{s.objective}

From the preceding we may summarize 
{\bf Born's statistical interpretation of the wave function} 
in the version of June 25, 1926 by the following five assertions
(expressed in modern terms):

(B1) A state of a quantum system at time $t$ is a stationary 
state, given by a simultaneous eigenstate of $H$, $\p$, and $J^2$.
(\sca{Bacciagaluppi} \cite{Bac2020} comments this in his footnote 9 
as follows: {\it ''This is not in fact stated explicitly but has to be 
read between the lines.''} The preceding subsection may thus be viewed 
as giving details on how to read between the lines.)

(B2) The corresponding state of motion is given by the joint 
eigenvalues $E=\hbar\omega$, $\hbar k$, and $\hbar^2j(j+1)$ of the 
commuting operators $H$, $\p$, and $J^2$. In the center of mass frame, 
$\p=0$. Then the quantum numbers are determined by $E$ and 
nondegeneracy. (Degenerate cases are excluded by assumption.)

(B3) Every quantum system is at each time $t$ in a unique 
(stationary) state, changing at random times by making a transition 
(a \bfix{quantum jump}). This is deemed obvious from spectroscopic 
experience.

(B4) Energy and momentum are exactly conserved. 

(B5) Associated with each quantum system is a guiding 
(de Broglie--Schr\"odinger) wave that determines through its spectral 
decomposition the probabililty of being in a particular state. 

Note the objective formulation of all statements, without any 
reference to measurement.

Born's interpretation met with strong resistance by Schr\"odinger, who
allegedly said -- as \sca{Heisenberg} \cite[p.6]{Hei1946} recalls much 
later -- at a conference in Kopenhagen in September 1926, 
{\it ''But should this damned quantum jumping persist, then I regret 
having even worked on this subject.''}
\\(German original: 
{\em''Wenn es doch bei dieser verdammten Quantenspringerei bleiben soll,
dann bedauere ich, dass ich mich \"uberhaupt mit diesem Gegenstand 
besch\"aftigt habe.''}.)
An exchange of letters between him and Born displays the two extreme 
poles between which the quantum community had to find a consensus in 
the subsequent years. 

Letter of Schr\"odinger to Born (November 2, 1926) 
(\sca{von Meyenn} \cite[p.329]{vMey}):
{\it ''I just skimmed (hence not yet really read) your last, just 
received, paper on the adiabatic theorem. [...]
But I still have the impression that you, and others who essentially 
share your view, are too deep under the spell of those concepts (such 
as stationary states, quantum jumps, etc.) that in the last twelve 
years established their right to be present in our thought, to give
full justice to the attempt to escape again form this thought pattern.
As an example I mention, e.g., that to the critique of the assumption 
that in the atom several eigen oscillations are simultaneously excited,
you ask the question, does it make sense to say the atom is 
simultaneously in several eigen oscillations? Aother remark seems to 
be to you a long way off: Always and everywhere else, where we have 
to do with oscillating systems, these generally oscillate not in the 
form of an  eigen oscillation but in a superposition of these.''}
\\(German original: 
{\it ''Ich habe eben Ihre letzte, ebenerhaltene Arbeit \"uber den 
Adiabatensatz \"uber\-flogen (also noch nicht wirklich gelesen). [...]
Ich habe aber doch den Eindruck, dass Sie und andere, die im 
Wesentlichen Ihre Ansicht teilen, zu tief im Banne derjenigen Begriffe 
stehen (wie station\"are Zust\"ande, Quantenspr\"unge usw.),
die sich in den letzten zw\"olf Jahren B\"urgerrecht in unserem Denken 
erworben haben, um dem Versuch, aus diesem Denkschema wieder 
herauszukommen, volle Gerechtigkeit widerfahren zu lassen. 
Als Beispiel f\"uhre ich z. B. an, dass Sie zur Kritik der Annahme: 
im Atom seien mehrere Eigenschwingungen gleichzeitig erregt,
die Frage stellen: hat es Sinn, zu sagen, das Atom befinde sich 
gleichzeitig in mehreren station\"aren Zust\"anden? Ganz ferne scheint 
Ihnen die andere naheliegende Bemerkung zu liegen: immer und \"uberall 
sonst, wo wir mit schwingungsf\"ahigen Systemen zu tun haben, schwingen 
dieselben im Allgemeinen nicht in der Form einer Eigenschwingung, 
sondern mit einem Gemisch derselben.''})

Letter of Born to Schr\"odinger (November 6, 1926): 
(\sca{von Meyenn} \cite[p.333f]{vMey}): 
{\it ''As most direct evidence of the stationary states we have the 
electron collisions, and here I think primarily of collisions of the 
second kind. We may taske, I think, as fact that an electron colliding 
with an atom in some particular state always gets only one increase 
in energy. Suppose now that in an atom several eigen oscillations can
 be simultaneously excited, how should one understand this? What is 
the use of your conter argument that the electron could itself be a 
'wave group'? For it is the atomic waves that correspond to discrete 
steps. Of course it is possible to say: The aton that oscillates in 
several frequencies responds in a special way, such that it transmits 
to the electron only the energy corresponding to one frequency. 
But this is only an excuse. It seems to me that Bohr's way of speaking 
is the natural description and summary of a large body of facts; 
therefore it is the task of every finer theory to justify this way of 
speaking.''}
\\(German original: 
{\it ''Als direkteste Evidenz der station\"aren 
Zust\"ande haben wir die Elektronenst\"osse, und hier denke ich vor 
allem an die St\"osse zweiter Art. Wir k\"onnen wohl als Tatsache 
nehmen, dass ein Elektron beim Zusammenstoss mit einem Atom, das in 
irgend einem Zustande ist, immer nur einen Energiezuwachs
bekommt. Wenn nun in einem Atom mehrere Eigenschwingungen gleichzeitig 
angeregt sein k\"onnten, wie sollte man das verstehen? Was n\"utzt da 
Ihr Gegenargument, dass das Elektron selber eine 'Wellengruppe' sein 
k\"onne? Es handelt sich doch um die Atomwellen, die diskreten Stufen 
entsprechen. Nat\"urlich ist es m\"oglich zu sagen: das in mehreren 
Frequenzen schwingende Atom reagiert eben in besonderer Weise so, dass 
es nur die einer Frequenz entsprechende Energie auf das Elektron 
\"ubertr\"agt. Aber das ist doch nur eine Ausrede. Mir scheint, dass 
die Bohrsche Redeweise die nat\"urliche Beschreibung und Zusammenfassung
eines grossen Tatsachenbereichs ist; darum ist es Aufgabe jeder 
feineren Theorie, diese Redeweise zu rechtfertigen.''})

In general, however, the statistical interpretation was quickly taken 
up by the quantum community. For example, on November 3, 1926,
\sca{Fowler} \cite{Fow1926} submitted a paper stating Born's 
interpretation: ''The state of the assembly is then specified of 
systems in the different quantum states, $n_1,n_2,\ldots,n_u,\ldots$.''
(p.434). Leter in the same paper, he speculates (on p.447) about the 
position probability density:
''We venture in conclusion some speculations on the subject of 
the statistical distribution of the particles in space. It is of some 
importance to remember that in the new statistics space distribution 
laws as such have almost ceased to exist, and this is probably in 
accord with the requirements of the new mechanics. The primary 
distribution laws are concerned only with the distribution over the
characteristics - that is, the energy. At the same time, there must be 
some means of deriving the average number of molecules 'present' in a 
given volume element.''

At that time, Pauli had already obtained the position probability 
density interpretation, but only in a letter to Heisenberg:

Letter of Pauli to Heisenberg, Oktober 19, 1926 
(\sca{Pauli} \cite[p.347]{Pau1979}):
{\it ''Now it is thus: all diagonal elements of the matrices (at least 
of functions of $p$ alone or of $q$ alone) one can interpret
kinematically  already now. For one can first ask for the probability 
that in a particular stationary state, the coordinates $q_k$ of the 
particles ($k=1,\ldots,f$) lie between $q_k$ und $q_k+dq_k$. The answer 
is $|\psi(q_1,\ldots,q_f)|^2dq_1\ldots dq_f$, if $\psi$ is 
Schr\"odinger's eigenfunction.''} 
\\(German original: 
{\it ''Nun ist es so: alle Diagonalelemente der Matrizen (wenigstens 
von Funktionen der $p$ allein oder der $q$ allein) kann man \"uberhaupt 
schon jetzt kinematisch deuten. Denn man kann ja zun\"achst nach der 
Wahrscheinlichkeit fragen, dass in einem bestimmten station\"aren
Zustand des Systems die Koordinaten $q_k$ seiner Teilchen 
($k=1,\ldots,f$) zwischen $q_k$ und $q_k+dq_k$ liegen. Die Antwort 
hierauf ist $|\psi(q_1,\ldots,q_f)|^2dq_1\ldots dq_f$, wenn $\psi$ die
 Schr\"odingersche Eigenfunktion ist.''})

The position probability density interpretation was published in the
next year in a footnote of 
\sca{Pauli} \cite[p.83,Footnote 1]{Pau1927}:
{\it ''We want to interpret this [...] function in the spirit of the
ghost field view of Born given in his book \cite{Bor1926a,Bor1926b}
as follows: $|\psi(q_1\ldots q_f)|^2dq_1\ldots dq_f$ is the probability 
that in the respective quantum state of the system these coordinates lie
simultaneously in the respective volume element $dq_1\ldots dq_f$ 
of position space.''} 
\\(German original: 
{\it ''Wir wollen diese [...] Funktion im Sinne der von Born in seiner
Stossmechanik [\cite{Bor1926a,Bor1926b}] vertretenen
Auffassung des ''Gespensterfeldes'' folgendermassen deuten: Es ist
$|\psi(q_1\ldots q_f)|^2dq_1\ldots dq_f$ die Wahrscheinlichkeit daf\"ur,
dass im betreffenden Quantenzustand des Systems diese Koordinaten
sich zugleich im betreffenden Volumenelement $dq_1\ldots dq_f$ des
Lageraums befinden.''})
Apart from its objective formulation (no reference to measurement),
this is a special case of the universal formulation (BR-US) of the 
Born rule.

The 1927 paper by \sca{von Neumann} \cite[p.45]{vNeu1927a} generalized
Pauli's statement to arbitrary selfadjoint operators, again stated as an
objective (i.e., measurement independent) interpretation. For discrete
energy spectra and their energy levels, we still read p.48:
{\it ''unquantized states are impossible''} 
\\(German original: 
{\it ''nicht gequantelte Zust\"ande sind unm\"oglich''}). 
Thus he follows Born's view that all physically possible states are 
stationary.

Like Born, both Jordan and von Neumann talk about objective
properties of the system independent of measurement. But unlike Born
who ties these properties to the stationary state representation in
which momentum and energy act diagonally, Pauli ties it to the
position representation, where position acts diagonally, and von Neumann
allows it for arbitrary systems of commuting selfadjoint operators.

From either Born's or Pauli's statement one can easily obtain the
basis-indepen\-dent objective expectation form (BR-OE) of the Born rule,
either for functions $A$ of stationary state labels, or for functions 
$A$ of position.

The first published statement of (BR-OE) seems to be the 1927 paper
by \sca{Landau} \cite{Landau1927}. The formulas (4a) and (5),
corresponding in modern notation to $\<A\>:=\tr\rho A$ in the mixed 
case and $\<A\>:=\psi^*A\psi$ in the pure case, $\rho=\psi\psi^*$,
are interpreted in Footnote 2 as probability mean 
(German original: {\it ''Wahrscheinlichkeitsmittelwert''}). 
Though he mentions that this rule is already known, earlier formulas 
in print for the mean always assumed a representation in which the 
operator is already diagonal. Again there is no reference to 
measurement.

%%%%%%%%%%%%%%%%%%%%%%%%%%%%%%%%%%%%%%%%%%%%%%%%%%%%%%%%%%%%%
\subsection{Paradoxes and measurement context in Born's rule}
\label{s.paradox}

Already in 1926, \sca{Pauli} \cite[p.347]{Pau1979} noticed the paradox 
that his objective position probability and Born's objective momentum 
probability interpretation following from a spectral interpretation 
of the objective expectation form (BR-OE) of the Born rule are mutually 
incompatible. He expressed it in the following words: 
{\it ''One can view the worls with the $p$-eye and one can view it 
with the $q$-eye, but if one wants to open both eyes at the same time,
one gets crazy''}; German original:
{\it ''Man kann die Welt mit dem $p$-Auge und man kann sie mit dem 
$q$-Auge ansehen, aber wenn man beide Augen zugleich aufmachen will, 
dann wird man irre''} 

The 1927 paper by \sca{Jordan} \cite[p.811]{Jor1927} cites  
\sca{Pauli} \cite{Pau1927} and  
extends Born's second assumption further to an objective, measurement
independent probability interpretation of inner products (probability
amplitudes) of eigenstates of two arbitrary operators in place of 
position and momentum. without being aware of the conceptual problem 
this objective view poses when applied to noncommuting operators.

In 1927, \sca{Weyl} \cite{Weyl1927}, still using objective language 
(p.9: ''hat ... die Gr\"osse'') and no reference to measurement, 
pointed out on p.2f that, due to noncommutativity and the resulting 
complementarity, Jordan's objectively interpreted formalism is 
mathematically defective and cannot extend to general operators. 
The reason is (though Weyl does not say this explicitly) that there is 
no joint probability density for noncommuting operators, hence not for 
position and momentum. In particular, Born's stationary state 
probability interpretation and Pauli's position probability density 
interpretation cannot both claim objective status.

Still in the same year, \sca{von Neumann} \cite{vNeu1927b} also 
notes (on p. 248) the problems resulting from noncommuting quantities 
that cannot be observed simultaneously, However, unlike Weyl, 
von Neumann was motivated by a consideration on p.247 of 
the measurement of values in an ensemble of systems, taking the 
expectation to be the ensemble mean of the measured values. 
Specialized to a uniform (''einheitlich'') ensemble of systems in the 
same completely known (pure) state $\psi$ of norm one he then finds on 
p.258 that $\rho=\psi\psi^*$ with some normalized state vector $\psi$, 
giving $\<X\>:=\psi^*X\psi$.
Note that a logical paradox can be deduced from (BR-OE) and noncommuting
quantities only when one assumes von Neumann's condition D,
which forces the measurements to be projective measurements and 
therefore allows one to construct probability densities.

In the present terminology, von Neumann's interpretation
of quantum expectation values is the measured expectation form (BR-ME) 
of the Born rule. This move eliminated the mathematical paradox noted 
by Pauli and Weyl, by relating the notion of the value of an observable 
(being in the $n$th state, or having position $q$) directly to 
measurement. Left was only the nonmathematical (hence less dangerous) 
paradox that position and momentum (or energy) are not measurable 
simultaneously to arbitrary precision -- a conundrum that already was 
(earlier that year) given a precise mathematical form in the 
uncertainty relation of \sca{Heisenberg} \cite{Hei1927}.

Later discussions of the statistical interpretation gradually switched 
from the objective versions of the Born rule to the measurement-based 
versions. As we have seen, de Broglie still maintained both variants 
in 1929. This transition process was apparently completed in 1930 
with the appearance of the very influential book by \sca{Dirac} 
\cite{Dir1}. 
A review of the book, by \sca{Lennard-Jones} \cite{LenJ1931}, states:
{\it ''The book contains Dr. Dirac's philosophy of the relation of 
theoretical and experimental physics. He believes that the main object 
of theory is to determine the possible results of an experiment, and 
to determine the probability that any one of these results will 
actually occur under given conditions. He regards it as quite 
unnecessary that any satisfying description of the whole course of the 
phenomena should be given. A mathematical machine is set up, and 
without asserting or believing that it is the same as Nature's machine, 
we put in data at one end and take out the results at the other. 
As long as these results tally with those of Nature, (with the same 
data or initial conditions) we regard the machine as a satisfying 
theory. But so soon as a result is discovered not reproduced by the 
machine, we proceed to modify the machine until it produces the new 
result as well.
\\
One might have hoped that the object of the theoretical physics was 
rather more ambitious than Dirac is willing to allow, and that the 
steady march forward of physics was taking us further and further 
forward to a knowledge of the nature of things. But the theoretical 
physicist, it would seem, must for ever abandon any hope of providing 
a satisfying description of the whole course of phenomena.''}
This is the view that today still dominates the philosophy of quantum 
mechanics.

%%%%%%%%%%%%%%%%%%%%%%
\subsection{Knowledge}\label{s.knowledge}

Von Neumann's change of perspective introduced the prominent reference 
to ''knowledge'' (\sca{von Neumann} \cite[p.247]{vNeu1927b}). It would 
be interesting to have a thorough historical analysis of the meaning 
and use of the term knowledge in quantum mechanics. In these days, 
this term was not understood in a subjective way, but as the objective 
(through thought experiments theoretically accessible) knowledge of 
what is real and in principle observable about the system, whether 
observed or not. 

However, the usage of the term ''knowledge'' created a tension that 
never since disappeared, and was verbalized most prominently by Einstein
throughout his life. In his obituary on Einstein, 
\sca{Puli} \cite{Pauli1959} wrote in 1959: 
{\it '' [...] even when I could not agree with Einstein's views. 
'Physics is after all the description of reality' he said to me, 
continuing, with a sarcastic glance in my direction 'or should I 
perhaps say physics is the description of what one merely imagines?'
This question clearly shows Einstein's concern that the objective 
character of physics might be lost through a theory of the type of 
quantum mechanics, in that as a consequence of its wider conception 
of the objectivity of an explanation of nature the difference between 
physical reality and dream or hallucination might become blurred.\\
The objectivity of physics is however fully ensured in quantum mechanics
in the following sense. Although in principle, according to the theory, 
it is in general only the statistics of series of experiments that is 
determined by laws, the observer is unable, even in the unpredictable 
single case, to influence the result of his observation -- as for 
example the response of a counter at a particular instant of time.
Further, personal qualities of the observer do not come into the 
theory in any way -- the observation can be made by objective 
registering apparatus, the results of which are objectively available 
for anyone's inspection.''} 
\\(German original:
{\it ''[...] auch wenn ich Einsteins Ansichten nicht zustimmen konnte. 
'Physik ist doch die Beschreibung des Wirklichen', sagte er zu mir und 
fuhr mit einem sarkastischen Blick auf mich fort: 'oder soll ich 
vielleicht sagen, Physik ist die Beschreibung dessen, was man sich 
bloss einbildet?' Diese Frage zeigt deutlich Einsteins Besorgnis, dass 
durch eine Theorie vom Typus der Quantenmechanik der objektive 
Charakter der Physik verloren gehen k\"onnte, indem durch deren weitere 
Fassung der Objektivit\"at einer Naturerkl\"arung der Unterschied der 
physikalischen Wirklichkeit von Traum oder Halluzination veschwommen 
werden k\"onnte.\\
Die Objektivit\"at der Physik ist in der Quantenmechanik jedoch im 
folgenden Sinn voll gewahrt. Obwohl nach der Theorie im Prinzip im 
allgemeinen nur die Statistik von Versuchsreihen gesetz\-m\"assig 
bestimmt ist, kann der Beobachter auch im nicht voraussagbaren 
Einzelfall das Resultat seiner Beobachtung -- wie zum Beispiel das 
Ansprechen eines Z\"ahlers in einem bestimmten Zeitmoment -- nicht 
beeinflussen.\\
Auch gehen pers\"onliche Eigenschaften des Beobachters in keiner 
Weise in die Theorie ein, viel\-mehr kann die Beobachtung durch 
objektive Registrierapparate erfolgen, deren Resultate allen zur 
Einsicht objektiv vorliegen.''})

Subjective knowledge (that Einstein hinted at with his question
{\it ''or should I perhaps say physics is the description of what one 
merely imagines?''}) seriously entered physics only two years earlier, 
in 1957, with the ''subjective statistical mechanics'' of 
\sca{Jaynes} \cite{Jay1957}, thus aggravating the tension.

It is easy to see that the amount of knowledge needed to apply 
correctly the maximum entropy principle must be an intrinsic property 
of the system modeled. For example, take as system a turbulent fluid. 
If one knows only the total energy and the volume, and applies the 
maximum entropy principle, one gets a Helmholtz ensemble describing a 
fluid in equilibrium, not the turbulent fluid. Thus in spite of having 
used valid knowledge, the maximum entropy principle gives completely 
wrong results. 

To correctly describe the latter in second quantization with the 
maximum entropy principle, it is necessary and sufficient that we 
apply it to an (approximate) knowledge of (at least) the quantum 
expectation values of the field operators for currents and densities 
of the fluid at all points in the support of the fluid. 

But for a turbulent fluid one cannot get this (very detailed and very
quickly changing) knowledge from observation, hence cannot know it 
in any meaningful sense! Instead, in all quantum derivations of fluid 
mechanical equations, one simply assumes that this knowledge 
(or rather the corresponding maximum entropy state) exists, and then 
draw consequences through statistical mechanics. 

Thus the knowledge required to fix the state in nonequilibrium 
statistical mechanics is \bfi{assumed knowledge}, to be checked by its 
consequences. This makes sense only if the state (and hence any quantum 
expectation value) assumed in the model is an intrinsic property of 
the fluid, at least to the extent that the model predictions match 
observation. 

Thus knowledge is encoded into a particular description, an objective 
property of the description used to model the system.
It is what the particular model claims to know, independent of who 
''knows'' or ''uses'' it. 
A change of knowledge is therefore just a change of the details in the 
model used to describe a particular system. In particular, the 
situation regarding knowledge is precisely the same as in classical 
mechanics, where the models also depend on this kind of knowledge 
(or prejudice, if the model is chosen based on insufficient knowledge).

Therefore, as Pauli had emphasized, nothing personal is implied in the 
term ''knowledge''. There is no more subjectivity in quantum 
statistical mechanics than in classical engineering physics: 
The only (and very restrictive) freedom a subject striving for maximal 
predictivity has is to choose a model correctly describing the system 
under study. Once the model is chosen, it determines the quality of 
the predictions, and, by comparison with corresponding observations, 
the quality of the knowledge built into the model. If the quality of 
this knowledge is insufficient, the model is objectively wrong, and 
was so even before the observations where made.

%%%%%%%%%%%%%%%%%%%%%
\subsection{Collapse (state reduction)}\label{s.collapse}

Tied to the measurement formulations of the Born rule is 
a controversial \bfix{collapse} (or \bfix{state reduction}) postulate 
about the state after a measurement. State reduction was introduced 
informally in \sca{Heisenberg} \cite[p.186]{Hei1927} 
({\it ''Every determination of position therefore reduces the wave 
packet again to its original size''}; 
German original: {\it ''Jede Ortsbestimmung reduziert also das 
Wellenpaket wieder auf seine urspr\"ungliehe Gr\"osse''})
and formally established by the authority of Dirac in several editions 
of his book:

{\it ''The state of the system after the observation must be an
eigenstate of [the observable] $\alpha$, since the result of a
measurement of $\alpha$ for this state must be a certainty.''}
(\sca{Dirac} \cite[p.49]{Dir1}, first edition, 1930)
\\
{\it ''Thus after the first measurement has been made, the system is in
an eigenstate of the dynamical variable $\xi$, the eigenvalue it
belongs to being equal to the result of the first measurement.
This conclusion must still hold if the second measurement is not
actually made. In this way we see that a measurement always causes the
system to jump into an eigenstate of the dynamical variable that is
being measured, the eigenvalue this eigenstate belongs to being equal
to the result of the measurement.''} 
(\sca{Dirac} \cite[p.36]{Dir3}, third edition, 1936)

Dirac's statement was discredited (rightly, but without much success) 
in 1958 by another authority, 
\sca{Landau \& Lifshitz} \cite[Section 7]{LL.3}, 
who remark that the state after the measurement is in general not an 
eigenstate, This is apparent in the modern notion of a quantum 
process (see, e.g., \sca{Chuang \& Nielsen} \cite{ChuN7} or
\sca{Mohseni} et al. \cite{MohRL}), which gives the correct 
post-measurement state for arbitrary POVM measurements.

In 2007, \sca{Schlosshauer} \cite{Schl.book} still takes the collapse 
({\it ''jump into an eigenstate''}) to be part of what he calls the
''standard interpretation'' of quantum mechanics. But he does not 
count it as part of the Born rule (p.35). 

It would be interesting to have a more thorough history of the 
collapse postulate and the extent of its validity.

Closely related to the collapse postulate is the so-called 
\bfix{eigenvalue-eigenstate link} (\sca{Fine} \cite{Fine}), thoroughly 
discussed by \sca{Gilton} \cite{Gil}), which asserts that 
{\it ''if a system has a definite value for $X$, 
then the system is in an eigenstate of $X$.''}

\newpage
%%%%%%%%%%%%%%%%%%%%%%%%%%%%%%%%%%%%%%%%%%%%%%%%%%%%%%%%%%%%%%%%%%%%%%%%
\section{The Born rule today}\label{s.today}

The most detailed, universal form of the Born rule is a very complex 
statement requiring considerable mathematical maturity to be fully 
understood. \sca{Schlosshauer \& Fine} \cite{SchlF} mention as 
desirable that a derivation of the Born rule gives 
{\it ''physical insight into the emergence of quantum probabilities 
and the Born rule''}.
It is indeed strange that such a complicated rule should be part
of the basis of quantum theory. 

For this and other reasons, there were a host of attempts to derive 
the Born rule (in some measured form) from more natural assumptions;
see \sca{Allahverdyan} et al. \cite[Chapter 2]{AllBN1} for a thorough 
discussion of the state of the art until 2013, and 
\sca{ Vaidman} \cite{Vai20} (2020) for a more recent survey. 

All attempts to derive the Born rule are doomed to failure if they do 
not define, in terms of the formal apparatus of quantum mechanics, the 
meaning of measurement, a concept appearing in the Born rule but not in 
the remaining basic rules of quantum mechanics. But measurement is 
usually introduced tacitly by invoking a seemingly self-evident 
consequence of the Born rule. For example, use of the maximum entropy 
principle in a derivation implies circularity since by the discussion 
in Subsection \ref{s.POVM}, its application silently invokes (BR-C). 

Another frequent mistake is to derive a probability distribution 
matching that of Born's rule, without any actual measurement (with 
definite results) being involved. Only performed calculations within 
the theoretical machinery are performed, and only handwaving is used 
to claim that these calculations have anything to do with measurement.
This was the case in early, decoherence-based 'derivations' now known
(\sca{Schlosshauer} \cite{Schl.book}) to lack an essential ingredient, 
the solution of the ''unique outcome problem''.

Therefore all derivations (known to me) found in the literature are
either circular or unrelated to measurement, and hence of questionable 
value. 

As examples, we consider three proposed derivations that appeared in
the last few years. 

\sca{Schonfeld} \cite{Scho} (2021) assumes in his elementary (but 
somewhat heuristic) 'derivation' of Born's rule on p.4 
the seemingly self-evident fact that the thermal randomness of 
subcritical droplet formation {\it ''exists independent of any 
measurement axioms, i.e., independent of whether anyone observes it.''} 
While his description gives some insight into the origin of the Born 
rule, the problem of deriving how random nucleation (rather than a 
mixture of all possible mucleations) arises from a quantum statistical 
model (cf. Subsection \ref{s.missing}) is not even touched. 
The traditional statistical account of nucleation by 
\sca{Langer} \cite{Lan69} is purely classical. A quantum derivation of 
nucleation was attempted, e.g., by \sca{Lombardo} et al. \cite{LomMR}. 
But (like many others) they assume without discussion that (under 
reasonable conditions) the {\it ''Wigner function can then be 
interpreted as a classical probability distribution for coordinates 
and momenta''}, an assertion that can be maintained as referring to a 
statistical distribution only by invoking the condensed version 
(BR-C) of the Born rule. 

\sca{Chanda \& Bhattacharyya} \cite{ChaB} (2021) claim on p.2 the 
emergence of Born's rule in {\it ''a dynamical model without an 
explicit invocation of an apparatus''}. Thus they fall into the second 
trap menitioned above, that nothing ever is measured in their analysis. 
(In addition, they assume in their derivation not the unitary dynamics 
of a closed system but the dissipative Lindblad dynamics, which would 
have to be derived without using Born's rule.)

\sca{Aharonov \& Shushi} \cite{AhaS} (2024) explicitly assume in their 
derivation -- in the line after (1) -- that measurement outcomes are 
eigenvalues, which is part of all spectral forms of the Born rule, and
therefore should be proved in a complete derivation. In the line after 
(9), an observed value comes out of the blue by simple handwaving.
(In addition, they explicitly assume a postulate, detailed after (12), 
which makes claims about unobservable properties of a particle between 
measurements -- without even spelling out what a property should be.
Thus instead of Born's rule with its clear operational meaning they 
have to postulate an obscure assumptions unlikely to be acceptable to 
many physicists.)

%%%%%%%%%%%%%%%%%%%%%%%%%%%%%%%%%%%%%%%%%%%%
\subsection{The detector response principle}\label{s.DRP}

In order not to follow into the trap of circularity, it is necessary 
to make the place explicit where measurement enters, and to define the 
properties a detector must have in order that its observations count as 
measurements. Therefore we begin with a precise conceptual framework 
within which some versions of the Born rule can be derived. 

Following \sca{Neumaier \& Westra} \cite[Part II]{NeuW}, we give
precise definitions for quantum expectations, statistical expectations, 
and quantum measurement devices, and derive from these the formulations 
(BR-POVM) and (BR-C) of the Born rule. 

We model a \bfix{quantum system} on the formal level using as 
\bfix{state space} $\Hz$ a Hermitian vector space (i.e., a dense 
subspace of a Hilbert space), and write $\Ez:=\Lin\Hz$ for the space 
of everywhere defined linear operators on $\Hz$. 
For single-particle states, $\Hz$ is the Schwartz space of smooth 
functions of position whose Fourier transform is also smooth. Then 
$\Ez$ is an operator algebra which contains the components of position 
$q$ and momentum $p$, and hence contains all normally ordered 
polynomials in $q$ and $p$. 

We characterize a \bfix{quantum source} by a positive semidefinite
Hermitian \bfix{density operator} (or \bfix{density matrix} if 
$\Hz=\Cz^n$) $\rho\in\Lin\Hz$. The \bfix{state} of the source is the 
positive linear mapping $\<\cdot\>$ that assigns to each $X\in \Lin \Hz$
its \bfi{quantum value}
\lbeq{e.qEx}
\<X\>:=\tr\rho X.
\eeq
As customary, we also refer to $\rho$ as the \bfix{state}, since states 
and density operators are in a 1-to-1 correspondence.
More generally, given a fixed state $\<\cdot\>$, the 
\bfix{quantum value} of a vector $X\in\Ez^m$ with operator components 
$X_j\in\Ez$ is the vector
\[
\ol X=\<X\>\in\Cz^m
\]
with components $\ol X_j=\<X_j\>$. 
Any property of the source relevant for quantum detection can be 
expressed as a function of quantum values. We call the number
\[
I:=\<1\>=\tr\rho
\]
the \bfix{intensity} of the source. The intensity is nonnegative since 
$\rho$ is positive semidefinite. $\rho=0$ defines the 
\bfix{empty state}; it is the only state with zero intensity. 
Note the slight difference compared to conventional density operators, 
where the trace is instead fixed to be one. 
(In contrast to our convention to normalize the trace of $\rho$ to the
intensity, it is customary in the literature to normalize the intensity 
to be $1$. This can be achieved by dividing $\rho$ and all quantum 
values by the intensity. A disatvantage of this normalization is that 
$\rho$ loses one degree of freedom and the intensity of the source is 
no longer represented by the state.)

A source and its state are called \bfix{pure} if the density operator
has rank 1, and hence is given by $\rho=\psi\psi^*$ for some
\bfix{state vector} $\psi$; if the source or state is not pure it is
called \bfix{mixed}. In a pure state, the quantum value takes the form
\[
\<X\>=\tr\psi\psi^* X =\psi^*X\psi.
\]
Since experimentally realizable detectors always produce only a finite 
number of possible results, we make the following definition.
A \bfix{quantum measurement device} (in the following just called a 
\bfix{measurement device}) is characterized by a collection of 
\bfix{detection elements} labelled by labels $k$ from a finite set $K$ 
satisfying the \bfix{detector response principle}, given by the 
following postulate, (In order that a measured mean rate has a sensible 
operational meaning, the source must be reasonably stationary, at least 
during the time measurements are taken.)

\bfix{(DRP)}:
{\it A detection element $k$ responds to an incident stationary
source with density operator $\rho$ with a nonnegative mean rate $p_k$ 
depending linearly on $\rho$. The mean rates sum to the intensity of 
the source. Each $p_k$ is positive for at least one density operator 
$\rho$.}
(A slightly less precise version is already in my unpublished 2019 
manuscript \cite[p.8]{Neu.BornM}.)

Unlike the Born rule, the DRP is not a postulate added to the formal 
core of quantum mechanics. Instead, it defines what it means for a 
physical (quantum) object to be a quantum detector. The DRP gives  
a simply property whose validity can be experimentally tested in 
concrete cases and decides whether or not a piece of matter may be 
regarded as a quantum detector. Thus it is fully compatible with the 
whole mathematical apparatus of quantum mechanics (including 
unitary evolution), without the need to 'shut up and calculate'.

The abundant existence of quantum measurement devices satisfying the DRP
is an extremely well established empirical fact. 
The immediate physical significance of the DRP makes it an excellent 
starting point for the statistical interpretation of quantum physics; 
in particular, it leads to a simple, transparent derivation of the 
Born rule. 

The DRP is valid whenever the discrete form (BR-DS) of the Born rule 
can be expected to hold, and allows us to give (in Subsection 
\ref{s.derived}) a straightforward derivation of the condensed form 
(BR-C) of the Born rule.

As shown in  \sca{Neumaier \& Westra} \cite[Part II]{NeuW}, the DRP 
leads naturally to all basic concepts and properties of modern 
quantum mechanics. In particular, it gives a precise operational 
meaning to quantum states, quantum detectors, quantum processes, and 
quantum instruments. This gives a perspective on the foundations of 
quantum mechanics that is quite different from the well-trodden path 
followed by most quantum mechanics textbooks.

%%%%%%%%%%%%%%%%%%%%%%%%%%%%%%%%%%%%%%%%%%%%%%%%%%%%%%%%%%%%%%%%
\subsection{The statistical interpretation of quantum mechanics}
\label{s.prob}

The key result for the theory of quantum measurements is the following 
\bfix{detector response theorem}, proved in \cite[Section 4.1.2]{NeuW}.

\begin{thm}\label{t.pPOVM}
For every measurement device, there is a unique discrete quantum
measure $P_k$ ($k\in K$) whose quantum values $\<P_k\>$ determine,
for every source with density operator $\rho$, the mean rates 
\lbeq{e.BornPOVM}
p_k=\<P_k\>=\tr\rho P_k  \for k\in K.
\eeq
\end{thm}

That the $P_k$ form a \bfix{discrete quantum measure} means that they 
are Hermitian, positive semidefinite, and sum up to the identity 
operator $1$. This is the natural quantum generalization of a discrete 
probability measure, a collection of nonnegative numbers that sum up 
to 1.

The detector response theorem characterizes the response of a quantum
measurement device in terms of a quantum measure.  
A quantum measurement device produces in the low intensity
case a stochastic sequence of \bfix{detection events}, but makes no
direct claims about values being measured. It just says which one of the
detection elements making up the measurement device responded at which 
time.

To go from detection events to measured numbers, one needs 
to provide in addition a \bfix{scale} that assigns to each 
detection element $k$ a real or complex number (or vector) $x_k$. 
We call the combination of a measurement device with a scale a 
\bfix{quantum detector}. We say that the detector \bfix{measures} the 
\bfix{quantity}
\lbeq{e.obs}
X:=\sum_{k\in K} x_kP_k.
\eeq
The same quantum measurement device, 
equipped with different scales, measures different quantities. 
When the density operator is normalized to intensity one the response 
rates $p_k$ form a discrete probability measure. In this case we refer 
to a response rate as response probability and to the quantum
value of $X$ as the quantum expectation of $X$. 

These values can be arbitrary numbers or vectors $x_k$ (or even more 
complex mathematical entities from a vector space) --
whatever has been written on the scale a pointer points to, or whatever
has been programmed to be written by an automatic digital recording
device.

A quantum detector may be considered as a technically precise version
of the informal notion of an \bfix{observer} that figures prominently in
the foundations of quantum mechanics. It removes from the latter term
all anthropomorphic connotations.

%%%%%%%%%%%%%%%%%%%%%%%%%%%%%%%%%%%%%%%%
\subsection{Derivation of the Born rule}\label{s.derived}

If the intensity is normalized to one then $\D\sum_k p_k=1$, so that 
the normalized mean rate $p_k$ can be interpreted as the 
\bfix{response probability} of detector element $k$ given some response,
or as the \bfix{detection probability} for the $k$th detection event. 
This gives an intuitive meaning for the $p_k$ in case of low intensity 
measurements. 
Formula \gzit{e.BornPOVM}, derived here from very simple first 
principles, becomes \gzit{e.rhoP}, but with $P_k$ not restricted to 
projections. Thus we have proved the POVM form (BR-POVM) of the Born 
rule. It gives the theoretical quantum values $\<P_k\>$ a statistical 
interpretation as response probabilities $p_k$ of a quantum measurement 
device. Note that all probabilities are classical probabilities in
the sense of \sca{Kolmogorov} \cite{Kol}, as used everywhere in 
statistics, and can be approximated by relative frequencies with an 
accuracy given by the law of large numbers. 

The results of a detector in a sequence of repeated events define a
random variable or random vector $x_k$ that allows us to define the 
\bfix{statistical expectation}
\lbeq{e.statEx}
\E(f(x_k)):=\sum_{k\in K} p_kf(x_k)
\eeq
of any function $f(x_k)$. 
As always in statistics, this statistical 
expectation is operationally approximated by finite sample means of 
$f(x)$, where $x$ ranges over a sequence of actually measured values. 
However, the exact statistical expectation is an 
abstraction of this; it works with a nonoperational probabilistic limit 
of infinitely many measured values, so that the replacement of relative 
frequencies in a sample by probabilities is justified. 

From \gzit{e.BornPOVM} and \gzit{e.obs}, we find the formula
\lbeq{e.BornEx}
\E(x_k)=\tr\rho X=\langle X\rangle
\eeq
for the statistical expectation of the measurement results $x_k$
obtained from a source with density operator $\rho$. Comparing with
\gzit{e.qEx}, we see that the statistical expectation of measurement
results coincides with the theoretical quantum value of $X$ evaluated
in the state $\rho$ of the source. This is the measured expectation 
form (BR-ME) of the Born rule. It gives the purely theoretical notion 
of a quantum value an operational statistical interpretation in terms 
of expectations of measurement results of a quantum detector. If we 
call the quantum value $\<X\>$ the \bfix{quantum expectation} of $X$, 
we find the Born rule \gzit{e.BornEx} in the condensed form (BR-C).

It is shown in \cite[Part II]{NeuW} that the DRP together with 
techniques from quantum tomography allow one to reconstruct from the 
DRP the complete edifice of quantum mechanics, including its dynamical 
and spectral consequences. 
In particular, the DRP is precisely what is needed to obtain the 
concepts and results from quantum information thoery. 

According to our discussion, the possible values obtained when 
measuring a particular quantity $X$ depend on the decomposition 
\gzit{e.obs} used to construct the scale.
That this decomposition is ambiguous follows from the fact 
that the scale is not determined by the quantity $X$ measured. Since 
the scale determines the measurement results, this means that one can 
with equal right ascribe different results to the measurement of the 
same quantity $X$. Thus different detectors measuring the same quantity 
$X$ may have different sets of possible measurement results. In other 
words, the same quantity $X$ can be measured by detectors with 
different mathematical characteristics, and in particular with 
different measurement results that need not have anything to do with 
the eigenvalues of $X$. In the terminology of Subsection 
\ref{s.collapse}, this means that the eigenstate-eigenvalue link is 
broken.

This is in full agreement with the standard recipes for drawing 
inferences from inaccurate measurement results. Just like in classical 
mechanics, they are emergent imperfections due to the experimental 
conditions in which the measurements are made. 
The situation is precisely the same as in classical metrology, where 
observable quantities always have a true value determined by the 
theoretical description, and all randomness in measurements is assumed
to be due to measurement noise.

\bigskip

We now connect our developments to the traditional
spectral view of quantum measurements.
We call a discrete quantum measure \bfix{projective} if the $P_k$
satisfy the \bfix{orthogonality relations}
\lbeq{e.orthP}
P_jP_k=\delta_{jk}P_k \for j,k\in K.
\eeq
We say that a detector measuring $X$ performs a
\bfix{projective measurement} of $X$ if its quantum measure is
projective.

In practice, the orthogonality relations \gzit{e.orthP} can often be
implemented only approximately (due to problems with efficiency, losses,
inaccurate preparation of directions, etc.). 
Thus projective measurements are unstable under imperfections in the
detector. Therefore measurements 
satisfying the spectral version of the Born rule (i.e., projective 
measurements) are almost always idealizations.
They are realistic only under special circumstances.
Examples are two detection elements behind
polarization filters perfectly polarizing in two orthogonal directions,
or the arrangement in an idealized Stern--Gerlach experiment for spin
measurement. Most measurements, and in particular all measurements of
quantities with a continuous spectrum, are not projective.

The orthogonality relations imply that $XP_k=x_kP_k$. Since the $P_k$
sum up to $1$, any $\psi\in\Hz$ can be decomposed into a sum
$\psi=\D\sum_k P_k\psi =\sum_k\psi_k$ of vectors $\psi_k:=P_k\psi$
satisfying the equation $X\psi_k=XP_k\psi=x_kP_k\psi=x_k\psi_k$.
Therefore each $\psi_k$ (if nonzero) is an eigenvector of $X$ (or of 
each component of $X$ in case the $x_k$ are not just numbers) 
corresponding to the eigenvalue $x_k$ of $X$. Since $P_k^2=P_k=P_k^*$, 
the $P_k$ are orthogonal projectors to the eigenspaces of the $x_k$.

When the $x_k$ are numbers,
this implies that $X$ is an operator with a finite spectrum. Moreover,
$X$ and $X^*$ commute, i.e., $X$ is a normal operator, and in case the
$x_k$ are real numbers, a Hermitian, self-adjoint operator. (This is the
setting traditionally assumed from the outset.) When the
$x_k$ are not numbers, our analysis implies that the components of $X$
are mutually commuting normal operators with a finite joint spectrum,
and if all $x_k$ have real components only, the components of $X$ are
Hermitian, self-adjoint operators.
Thus the projective setting is quite limited with respect to the
kind of quantities that it can represent.

For projective measurements, \gzit{e.obs} implies
\[
f(X^*,X)=\sum_k f(\ol x_k,x_k) P_k
\]
for all functions $f$ for which the right hand side is defined.
Therefore the modified scale $f(x^*,x)$ measures $f(X^*,X)$, as we
are accustomed from classical measurements, and defines a projective
measurement of it. But when the components of $X$ are not normal or do
not commute, this relation does not hold.

Thus we recover the traditional spectral setting as a consequence of 
the general approach, restricted to the special case where the 
components of $X$ are commuting self-adjoint (or at least normal) 
operators, hence have a joint spectral resolution with real (or complex)
eigenvalues $x_k$, and the $P_k$ are the projection operators to the 
eigenspaces of $X$.

%%%%%%%%%%%%%%%%%%%%%%%%%%%%%%%%%%%%%%%%%%%%%%%%%%%%%%%%%%%%%%%%%
\subsection{The domain of validity of various forms of Born rule}
\label{s.valid}

It is well established that quantum mechanics applies universally on 
all physical scales, not only in the microscopic domain. This implies 
that the Born rule cannot apply to arbitrary physical measurements. 
It does not cover the multitude of situations where typically only a 
single measurement of a observable quantity is made. In particular, 
the Born rule does not apply to typical macroscopic measurements, 
whose essentially deterministic predictions are derived from quantum 
statistical mechanics.

Many other things physicists measure have no simple interpretation in 
terms of the Born rule. Often lots of approximate computations are 
involved before raw observations lead to measurement results. Other
examples include spectral lines and widths, life times of unstable 
particles, scattering cross sections, or chemical reaction rates. 

Thus we now address the task of describing the domain of validity of 
the various formulations of the Born rule, delineating the class of 
measurements for which they gives a valid description.

%%%%%%%%%%%%%%%%%%%%%%%%%%%%%%%%%%%%%%%%%%%%%%%%%
\subsubsection{Invalidity of the objective forms}

As discussed in Subsection \ref{s.paradox}, the objective forms 
(BR-OSc) and (BR-OE) of the Born rule are untenable, since the lack 
of commutativity of the operator algebra leads to the impossibility to
assign a joint probability distribution to position and momentum 
(and to other noncommyting observables). 

\sca{Peskin \& Schroeder} \cite{PesS} appeal on p.104 to the 
objective expectation form (BR-OE) of the Born rule by considering 
{\it ''the probability for the initial state to scatter and become a 
final state of $n$ particles whose momenta lie in a small region''}.
But this is the sloppiness that adherents of 
{\it shut-up-and-calculate} allow themselves, and indeed, they refer 
to measurement on the page before.

%%%%%%%%%%%%%%%%%%%%%%%%%%%%%%%%%%%%%%%%%%%%%%%%%%%%%%%%%%%%%%%%%%
\subsubsection{Domain of validity of the measured scattering form}

\sca{Born} \cite{Bor1926b} mentioned already in 1926 that the objective 
scattering form (BR-OSc) of the Born rule is applicable only in the 
absence of degeneracy. The same holdes for the measured scattering form 
(BR-MSc).
Under this restriction the latter turned out to be impeccable even 
today. The additional rule that in the degenerate case one has to sum or
integrate over a complete basis of out-states in the eigenspace of the 
joint spectrum of the asymptotically conserved quantities allow 
the correct interpretation of arbitrary scattering processes. 

The measured scattering form (BR-MSc) of the Born rule is impeccable 
and remains until today the basis of the interpretation of S-matrix 
elements calculated in quantum mechanics and quantum field theory. 

The quantum field theory book by \sca{Weinberg} \cite{Wei.I} pays in 
(2.1.7) on p.50 lip service to the universal form (BR-US) of Born's 
rule. But the only place in the book where Born's rule is actually used
is on p.135: 
({\it ''The probability of a multi-particle system, which is in a state 
$\alpha$ before the interaction is turned on, isfound in a state $\beta$
after the interaction is turned off, is''} his formula (3.4.7). 
The wording ''is found' shows that the measured scattering form 
(BR-MSc) is employed to get the formula (3.4.11) for the transition 
rates in scattering processes. (Indeed, quantum field theory always 
predicts transition {\it rates}, no transition probabilities.) 
In this context it is interestiung to note how Weinberg sweeps 
(on p.134) the associated measurement problem und er the carpet:
({\it ''with the excuse that (as far as I know) no interesting open 
problems in physics hinge on getting the fine points right regarding 
these matters''}.

%%%%%%%%%%%%%%%%%%%%%%%%%%%%%%%%%%%%%%%%%%%%%%%%%%%%%%%%%%%%%%%%%%%%%%
\subsubsection{Domain of validity of the finite and the discrete form}

On the other hand, the finite form (BR-FS) of the Born rule needs 
four conditions for its validity. It is valid precisely for measuring
quantities\\
1.
with only discrete spectrum,\\
2.
measured over and over again in identical states (to make sense of the
probabilities), where\\
3.
the difference of adjacent eigenvalues is significantly larger than
the measurement resolution, and where\\
4.
the measured value is adjusted to exactly match the spectrum, which
must be known exactly prior to the measurement.

These conditions are satisfied for measurements in the form of clicks, 
flashes or events (particle tracks) in scattering experiments, and 
perhaps only then.

That these requirements are needed is due to the insistence of the 
spectral formulations that only exact eigenvalues are measured, 
a condition violated in many quantum measurement situations.

\pt
A measurement of the mass of a relativistic particle with 4-momentum 
$p$ never yield an exact eigenvalue of the mass operator 
$M:=\sqrt{p^2}$. Indeed, the masses of most particles are only 
inaccurately known.

\pt
Energy measurements of a system at energies below the dissociation 
threshold (i.e., where the spectrum of the Hamiltonian $H$, the 
associated quantity, is discrete), almost never yield an exact 
eigenvalue of $H$, as the discrete form of the Born rule requires. 
Indeed, the energy levels of most realistic quantum systems are only 
inaccurately known. For example, nobody knows the exact value of the 
Lamb shift, a difference of eigenvalues of the Hamiltonian of the 
hydrogen atom; the first reasonably precise measurement was even worth 
a Nobel prize (1955 for Willis Lamb). 

\pt
The discrete part of the spectrum of a composite system is usually very 
narrowly spaced and precise energy levels are known only for the 
simplest systems in the simplest approximations. Thus the Born rule 
does not apply to the total energy of a composite system, according to 
Dirac one of the key quantities in quantum physics. In particular, the 
Born rule cannot be used to justify the canonical ensemble formalism 
of statistical mechanics; it can at best motivate it.

\pt
Real measurements usually produce numbers that are themselves subject 
to uncertainty (\sca{NIST} \cite{SI2}), and rarely the exact numbers 
that the discrete form (BR-DS) of the Born rule requires.
This implies that the spectral form of the Born rule paints an 
inadequate, idealized picture whenever eigenvalues are only 
approximately known and must therefore be inferred experimentally. 
For example, a \bfix{Stern--Gerlach experiment} measures (according to 
the common 
textbook story) the eigenvalues of the spin operator 
$L_3=\hbar\sigma_3$ (with angular momentum units), where $\hbar$ 
is Planck's constant. The eigenvalues are $\pm\hbar/2$. 
Taking the spectral form of his rule literally, Born could have deduced 
in 1927 from the Stern--Gerlach experiment the exact value of Planck's 
constant! But the original Stern--Gerlach experiment produced on the 
screen only two overlapping lips of silver, from which one cannot get 
an accurate value for $\hbar/2$. Indeed,
\sca{Busch} et al. \cite[Example 1, p.7]{BusGL} write:
{\it The following 'laboratory report' of the historic Stern-Gerlach
experiment stands quite in contrast to the usual textbook 'caricatures'.
A beam of silver atoms, produced in a furnace, is directed through an
inhomogeneous magnetic field, eventually impinging on a glass plate.
[...]
Only visual measurements through a microscope were made. No statistics
on the distributions were made, nor did one obtain 'two spots' as is
stated in some texts. The beam was clearly split into distinguishable
but not disjoint beams.}

%%%%%%%%%%%%%%%%%%%%%%%%%%%%%%%%%%%%%%%%%%%%%%%%%%%%%%%%
\subsubsection{Domain of validity of the universal form}

The universal form of the (BR-US) Born rule inherits all limitations
of the discrete form, but has additional limitations.

\pt
The joint measurements of quantities represented by operators that 
do not commute cannot even be formulated in the textbook setting 
of projective measurements. Thus the Born rule in its universal form 
does not apply, and one needs a quantum measure (POVM) that is not 
projective to model the measurement. 
For example, this applies for a simultaneous low resolution 
measurement of position and momentum by inferring it from the trace of 
a particle in a cloud chamber.

\pt
As first observed by \sca{Heisenberg} \cite[p.25]{Hei1930}, 
the Born rule implies a tiny but positive probability that an 
electron bound to an atom is detected light years away from the atom:
{\it ''The result is more remarkable that it seems at first. For it 
is known that $\psi^*\psi$ decreases exponentially with increasing 
distance. Hence there is always a positive probability for finding 
the electron very far from the nucleus.''} 
\\(German original: 
{\it ''Das Resultat ist aber merkw\"urdiger, als es im ersten 
Augenblick den Anschein hat. Bekanntlich nimmt $\psi^*\psi$ 
exponentiell mit wachsendem Abstand vom Atomkern ab. Also besteht immer 
noch eine endliche Wahrscheinlichkeit daf\"ur, das Elektron in sehr 
weitem Abstand vom Atomkern zu finden.''})
Therefore $|\psi(x)|^2$ is unlikely to be the {\it exact} probability 
density for being detected at $x$, as (BR-US) would require.
This indicates that for observables with continuous spectrum, (BR-US) 
must be an idealization.

%%%%%%%%%%%%%%%%%%%%%%%%%%%%%%%%%%%%%%%%%%%%%%%%%%%%%%%%
\subsubsection{Domain of validity of the condensed form}

The statistical interpretation based on the Born rule in the condensed
form (BR-C) and in the POVM form (BR-POVM) are derivable from the DRP,
and give a correct account of the actual experimental situation. 
They apply without idealization to results of arbitrary quantum 
measurements in the sense of the DRP. For a large number of examples 
see \sca{Neumaier \& Westra} \cite[Section 6.2]{NeuW}.

However, (BR-C) and (BR-POVM) still have the same limitations as the 
DRP, and cannot be applied to measurement of the kind discussed at the 
beginning of this subsection, such as macroscopic measurements or the 
measurements of spectral lines. 
For example, spectroscopy-based high precision measurements such as 
the gyromagnetic ratio of the electron are not quantum measurements 
but measurements of numerical parameters in specific quantum models; 
see the detailed discussion in 
\sca{Neumaier \& Westra} \cite[Section 5.2]{NeuW}.

%%%%%%%%%%%%%%%%%%%%%%%%%%%%%%%%%%%%%%%%%%%%%%%%
\subsection{What is missing in the foundations?}\label{s.missing}

At every steadily progressing now, Nature somehow produces from 
{\it what has been up to now} (defined by what is observable, at least 
in principle) that {\it what has been up to a while later}.  

Physics is supposed to quantitatively describe and understand this
in terms of physical theories and models, with which predictions can be 
made, and in terms of experiments that evaluate some of the details 
of {\it what has been up to now}, with which such predictions can be 
prompted, verified, or refuted. But Nature proceeds without caring 
about our physical models and the knowledge built into these, and has 
done so long before the first living observers existed.

The fundamental physical state $\lambda(t)$ is a complete description 
of the (to us unknown) collection of everything that {\it has been up 
to time $t$}. The $\lambda(t)$ for all $t$ fully describe 
{\it what is}. This defines an effective ontology based on what is 
observable in principle.

The (to us unknown) fundamental dynamical law of physics 
determines how $\lambda(t)$ for $t>t_0$ is related to $\lambda(t_0)$.

Our physical models and the knowledge built into these reflect our 
collective attempts to distill from {\it what has been up to now} an 
approximate picture of the fundamental physical state and the 
fundamental dynamical law that allows us to predict a good 
approximation to {\it what is}.

Curiously, at present, our physical models are not in terms of what is
(i.e., of approximations to $\lambda(t)$, part of which is observed) 
but in terms of states $\psi$ or $\rho$ and observables $X$ only 
indirectly related to what is.
This indirect relation between the quantum theory of states and 
observables and reality is is given by one of the versions of the Born 
rule.

The above notion of {\it what is} must be distingushed from the 
theoretical ontology of certain interpretations of quantum mechanics: 
In the Copenhagen interpretation, {\it what is} is the classical world,
delineated from the quantum world by a movable Heisenberg cut. 
In the consistent histories approach (\sca{Griffiths} \cite{Gri}),
{\it what is} is the single history up to our present, which is just 
one of an uncountable collection of consistent histories about which 
theory only contributes uncheckable probability assignments to each of 
them. 
In many world interpretations, a preferred ontological status is 
given to a postulated universal wave function, a theoretical figment of 
abstraction, whereas {\it what is} (in the present sense) is degraded 
to a particular world amidst infinitely many worlds, namely the single 
distinguished world that is our actual past, and from which we have to 
learn everything we can know about Nature, independent of any weight 
one might give to this world.
In Bohmian mechanics, infinitely precise positions and the exact wave 
function are declared to have a preferred ontological status 
-- a poor, theoretical substitute for the colorful {\it what is} that 
we experience.
In QBism (\sca{Fuchs} \cite[Section 3]{Fuc23}), {\it what is} (in the 
present sense) is degraded to the world of a metaphysical society of
agents governed by moral imperatives of how they should bet.
In each case, nothing real (in the sense of our {\it what is}) is in 
the quantum theory (of states and observables) itself, a state of 
affairs criticized by Einstein throughout his life.

The Born rule depends on a notion of measurement that has no formal 
counterpart in quantum mechanical models. This means that we are 
currently unable to explain theoretically which formal quantum models 
of pieces of matter describe a quantum detector. The theoretical 
explanation for why certain objects are quantum detectors constitutes 
the so-called \bfi{quantum measurement problem}. 

If we accept the DRP, the problem is to give mathematical conditions 
on certain individual quantum objects described by microscopic quantum 
mechanics that guarantee that the DRP holds for a suitable mathemetcal 
definition of what it means for the quantum model to respond at a 
particular time. This is a nontrivial unsolved problem in the 
statistical mechanics of detector properties.

In 1962, \sca{Daneri} et al. \cite{DanLP} initiated the discuss of 
quantum detectors by means of quantum statistical mechanics. However,
they assume the objective scattering form (BR-Sc) of the Born rule:
{\it ''the intermediate state (depending on the direction in which the 
particle is scattered) is the state in which ions are created inside 
the $k$-th counter''} (p.308). In addition, they assume that the 
detector is dissipative (p.307). The latter (but not some version of 
the Born rule) could be avoided by using a metastable quantum device 
coupled to a heat bath with a continuous frequency spectrum, along the 
lines of \sca{Caldeira \& Leggett} \cite{CalL83}.

The work by \sca{Allahverdyan} et al. \cite{AllBN1} goes some way 
towards resolving the problem. But they have to equate at some points 
the formal probabilities of statistical mechanics 
(cf. Subsection \ref{s.qEx}) with measured probabilities, thus silently 
invoking the Born rule. This is done at the beginning of Section 11, 
where they say that {\it ''the statistical interpretation of quantum 
mechanics emphasizes the idea that this theory, whether it deals with 
pure or mixed states, does not govern individual systems but 
statistical ensembles of systems''}. But it describes statistical 
ensembles only if the formal probabilities are taken to be observed
frequentist probabilities, which requires Born's rule. Thus the 
derivation is circular. The problem is still present in their recent 
paper \cite[p.254]{AllBN24}, since they assume the maximum entropy 
principle, which (as discussed above in Subsection \ref{s.qEx}) 
requires the condensed form (BR-C) of Born's rule. Moreover, to go
from an ensemble to an individual case, they need an additional
assumption called Postulate P. This postulate, {\it ''although 
intuitive, is not a consequence of the mere quantum principles and has 
a macroscopic nature. We will accept its inclusion in quantum theory 
as a postulate (motivated by the macroscopic size of M)''}.  
In effect they show that (BR-C) assumed for quantum statistical 
mechanics, together with their new Postulate P, implies that the 
measurement device they analyzed responds as the Born rule requires. 
Thus they reduced the measurement problem to their Postulate P. 

The decoherent history approach of \sca{Gell-Mann \& Hartle} \cite{GelH}
for the derivation of classical probability suffers from the problem of 
having to reinterpret formal smeared Wigner functions as classical 
probability densities (cf. my introduction to Section \ref{s.today}), 
whose statistical interpretation needs (BR-C).
Indeed, they assert on p.3378 that something is missing:
{\it ''we may begin to tackle the deep problem of introducing 
individuality into quantum mechanics. Actual alternative histories 
deal, of course, in large part with individual objects such as our 
galaxy, the Sun, the Earth, biological organisms on the Earth, and so 
forth. Yet discussions of quantum mechanics up to now have typically
treated such individual objects only as external systems,
labeled as 'observers' and 'pieces of apparatus'.''}

The main problem behind these difficulties is the need to find a 
rigorous derivation of a dissipative classical dynamics for a single 
pointer (or switch) without a silent recourse to the Born rule. 
A customary way to proceed (following \sca{M\"ohring \& Smilansky} 
\cite{MoeS}) is to start with the influence functional approach of 
\sca{Feynman \& Vernon} \cite{FeyV}. However, this approach has the 
Born rule built in -- namely into the derivation (p.122) of the basic 
functional integral representation of the transition probabilities. 

A solution of the measurement problem is indeed impossible unless one 
has a precise mathematical definition of when and how a mathematical 
model of a single quantum detector produces a definite 
response. E.g., \sca{Busch \& Lahti} \cite[p.375]{BusL3} write: 
{\it ''There is no consistent quantum measurement theory, 
unless an appropriate reinterpretation of what it means for an 
observable to have a definite value can be found.''}
This is the reason why, 100 years after the inception of quantum 
mechanics in terms of states and observables, the quantum measurement 
problem is still unsolved. It is the missing piece in the foundations 
of quantum mechanics.

\newpage
%%%%%%%%%%%%%%%%%%%%%%%%%%%%%%%%%%%%%%%%%%%%%%%%%%%%%%%%%%%%%%%%%%%%%

%%%%%%%%%%%%%%%%%%%%%%%%%%%%%%%%%%%%%%%%%%%%%%%%%%%%%%%%%%%%%%%%%%%%%%%%
\end{document}